\documentclass[%
 reprint,
 superscriptaddress,
 showpacs,preprintnumbers,
 amsmath,amssymb,
 aps,
]{revtex4-1}

\DeclareMathAlphabet{\mathsfit}{T1}{\sfdefault}{\mddefault}{\sldefault}
\SetMathAlphabet{\mathsfit}{bold}{T1}{\sfdefault}{\bfdefault}{\sldefault}

\usepackage{graphicx}
\graphicspath{{fig/}}   
\usepackage{dcolumn}
\usepackage{bm}
\usepackage{makecell}
\usepackage{braket} 
\usepackage{siunitx}
\usepackage{color}
\usepackage{xr}
\usepackage{upgreek}
\usepackage[colorlinks]{hyperref}
\usepackage[capitalize]{cleveref}
\usepackage[caption=false]{subfig}
\usepackage{multirow} 

\newcommand{\otoc}{\mathsfit{F}}
\newcommand{\comm}{\mathsfit{C}}
\newcommand{\fidelity}{\mathcal{F}}
\renewcommand{\Re}{\mathrm{Re}}
\newcommand{\tr}{\mathrm{Tr}}
\DeclareMathSymbol{\shortminus}{\mathbin}{AMSa}{"39}
\newcommand{\wq}{\omega_{\mathrm q}}
\newcommand{\SZ}{\hat\Sigma _z}
\newcommand*{\I}{\mathrm{i}}

\newcommand{\pr}{^{\prime}}

\begin{document}

\title{Probing quantum information propagation with out-of-time-ordered correlators}

\def\RLEaffil{Research Laboratory of Electronics, Massachusetts Institute of Technology, Cambridge, MA 02139, USA}
\def\LLaffil{MIT Lincoln Laboratory, Lexington, MA 02421, USA}
\def\Physaffil{Department of Physics, Massachusetts Institute of Technology, Cambridge, MA 02139, USA}
\def\EECSaffil{Department of Electrical Engineering and Computer Science, Massachusetts Institute of Technology, Cambridge, MA 02139, USA}
\def\Maryaffil{Laboratory for Physical Sciences, 8050 Greenmead Dr., College Park, MD 20740, USA}

\author{Jochen~Braum\"uller}
\email{jbraum@mit.edu}
\affiliation{\RLEaffil}
\author{Amir~H.~Karamlou}
\affiliation{\RLEaffil}
\affiliation{\EECSaffil}
\author{Yariv~Yanay}
\affiliation{\Maryaffil}
\author{Bharath~Kannan}
\affiliation{\RLEaffil}
\affiliation{\EECSaffil}
\author{David~Kim}
\affiliation{\LLaffil}
\author{Morten~Kjaergaard}
\affiliation{\RLEaffil}
\author{Alexander~Melville}
\author{Bethany~M.~Niedzielski}
\affiliation{\LLaffil}
\author{Youngkyu~Sung}
\affiliation{\RLEaffil}
\affiliation{\EECSaffil}
\author{Antti~Veps\"al\"ainen}
\author{Roni~Winik}
\affiliation{\RLEaffil}
\author{Jonilyn~L.~Yoder}
\affiliation{\LLaffil}
\author{Terry~P.~Orlando}
\affiliation{\RLEaffil}
\affiliation{\EECSaffil}
\author{Simon~Gustavsson}
\affiliation{\RLEaffil}
\author{Charles~Tahan}
\affiliation{\Maryaffil}
\author{William~D.~Oliver}
\affiliation{\RLEaffil}
\affiliation{\EECSaffil}
\affiliation{\LLaffil}
\affiliation{\Physaffil}

\date{\today}

\begin{abstract}

Interacting many-body quantum systems show a rich array of physical phenomena and dynamical properties, but are notoriously difficult to study: they are challenging analytically and exponentially difficult to simulate on classical computers. Small-scale quantum information processors hold the promise to efficiently emulate these systems, but characterizing their dynamics is experimentally challenging, requiring probes beyond simple correlation functions and multi-body tomographic methods. Here, we demonstrate the measurement of out-of-time-ordered correlators (OTOCs), one of the most effective tools for studying quantum system evolution and processes like quantum thermalization. We implement a 3x3 two-dimensional hard-core Bose-Hubbard lattice with a superconducting circuit, study its time-reversibility by performing a Loschmidt echo, and measure OTOCs that enable us to observe the propagation of quantum information. A central requirement for our experiments is the ability to coherently reverse time evolution, which we achieve with a digital-analog simulation scheme. In the presence of frequency disorder, we observe that localization can partially be overcome with more particles present, a possible signature of many-body localization in two dimensions.

\end{abstract}

\maketitle

\section{Introduction}

\begin{figure*}
\subfloat{\label{fig:timeevolution}}
\subfloat{\label{fig:manhattanschematic}}
\subfloat{\label{fig:chipphoto}}
\subfloat{\label{fig:timerevdiagram}}
\includegraphics{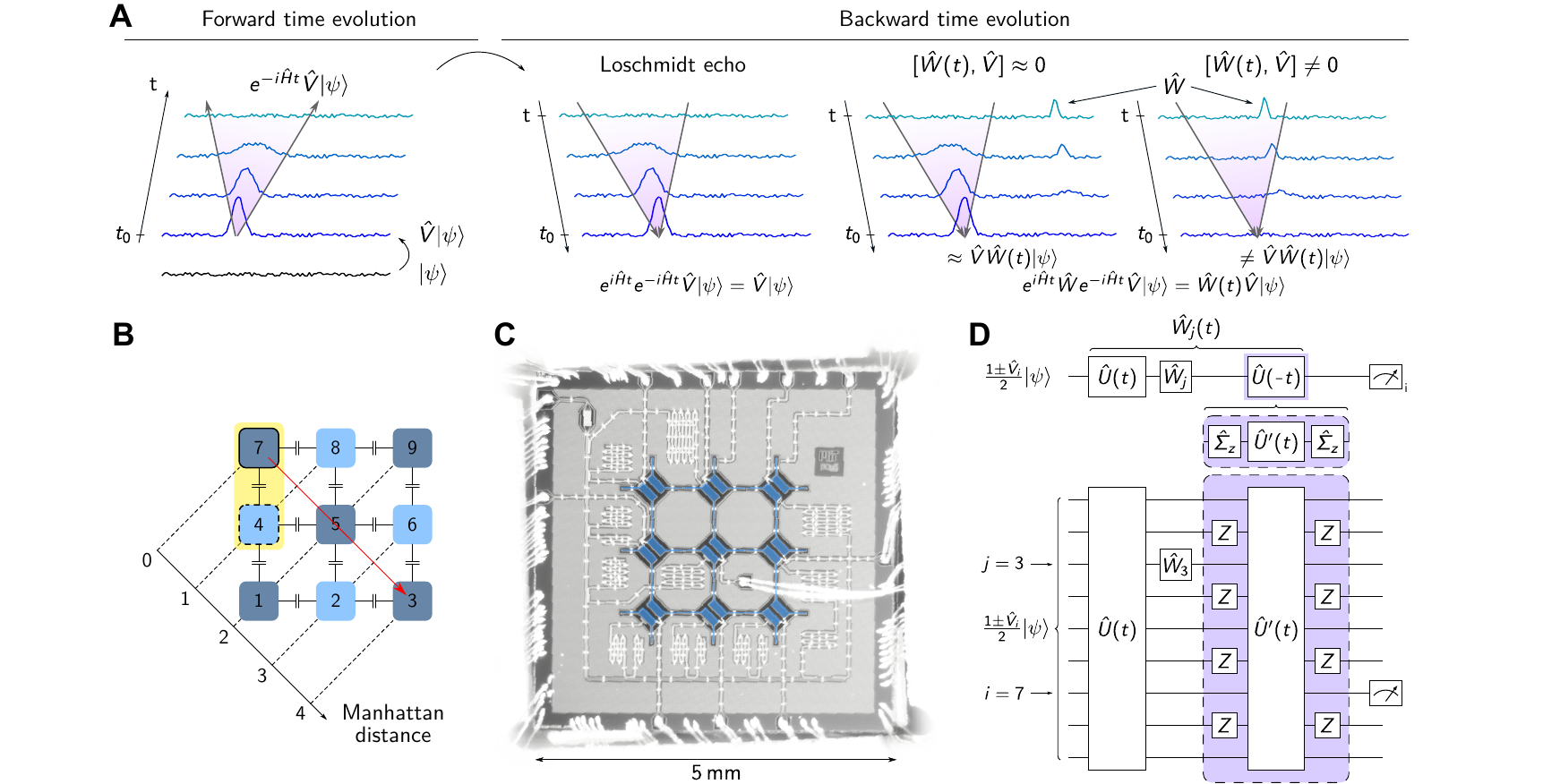}
\caption{
\textbf{Experimental concept}
(\textbf{A}) Schematic time evolution of a quantum system. From left to right: information about some initial perturbation $\hat V$ to the equilibrium state is scrambled into non-local lattice degrees of freedom. The original state $\hat V\Ket\psi$ re-materializes in a Loschmidt echo, after reversing time evolution. If the system is additionally perturbed by $\hat W$ prior to rewinding time, temporal reversibility is impeded if $\hat W$ falls within the light cone of $\hat V$ (purple), captured by the squared commutator $\comm$. The horizontal axis in the schematic represents some abstract spatial coordinate.
(\textbf{B}) Schematic device diagram of the 2d lattice used in our experiment. To study system dynamics, we use the indicated Manhattan distance (norm-1), which is the minimum number of sites traversed by a particle moving between lattice sites.
(\textbf{C}) Photograph of the superconducting circuit used in experiment. The capacitor pads of the transmon qubits are highlighted in blue.
(\textbf{D}) Digital-analog circuit diagram used to experimentally access OTOCs.
}
\label{fig:fig1}
\end{figure*}

Interacting quantum systems in the absence of strong disorder obey the eigenstate thermalization hypothesis: the system evolves towards a local thermal equilibrium state~\cite{Deutsch1991,Srednicki1994,Rigol2008,Neill2016}. During this process, information about any initial local perturbation is dispersed throughout the non-local degrees of freedom of the entire system~\cite{Neill2016,Blok2021,Mi2021}, a process referred to as information scrambling~\cite{Landsman2019,Joshi2020}. For a sufficiently long time evolution $\hat U(t)$, the information about an initial perturbation $\hat V$ is seemingly lost when probing the resulting state with local measurements (Fig.~\ref{fig:fig1}A, left panel). One way to retrieve the information hidden in the non-local degrees of freedom is to rewind the time evolution, such that the initial perturbation re-materializes, effectively reviving the original state $\hat U(\shortminus t)\hat U(t)\hat V\Ket\psi=\hat V\Ket\psi$. This sequence of consecutive forward and backward time evolution steps is called a Loschmidt echo~\cite{Peres1984,Yan2020} (Fig.~\ref{fig:fig1}, second panel).

An extension of the Loschmidt echo sequence can be used to study the propagation of quantum information in the system. As the system evolves, the perturbation $\hat V$ spreads into the lattice, with its `lattice reach' given by a light cone, signifying the portion of the system in causal connection with $\hat V$~\cite{Swingle2018}. This is the quantum analog of the light cone in general relativity, defining the speed of light. If the system is additionally perturbed at time $t$ (prior to the time reversal step) by a local operator $\hat W$, the reversibility of the time evolution is disturbed if $\hat W$ falls within the light cone (Fig.~\ref{fig:fig1}A, right panels). For two Hermitian, unitary local operators $\hat V$, $\hat W$, this operator growth -- the spreading of the operator's impact to an increasing number of lattice sites -- is quantified by the expectation value of the squared commutator~\cite{Swingle2018,Campisi2017}
\begin{equation}
\comm(t) =\langle[\hat W (t),\hat V]^\dagger [\hat W (t),\hat V]\rangle =2-2\Re\left[\otoc(t)\right],
\label{eq:otocdef}
\end{equation}
where $\otoc(t)=\langle \hat W(t)\hat V\hat W(t)\hat V\rangle$ is called the `out-of-time-ordered correlator' (OTOC). Intuitively, the OTOC measures the extent to which information about $\hat V$ has reached the lattice site where the perturbation $\hat W$ is applied at time $t$; if it is within the light cone indicated in Fig.~\ref{fig:fig1}A, the commutator is non-zero, while it approximately vanishes when $\hat W$ is outside the light cone. The OTOC cannot be obtained from a measurement of local operators, since the time arguments in $\otoc(t)$ are not time-ordered. Moreover, time-ordered correlators such as $\langle\hat W(t)\hat V\rangle$ decay rapidly in time for thermalizing systems and are therefore not suitable for characterizing operator growth~\cite{Swingle2018}.

We study the propagation of information in a strongly interacting two-dimensional (2d) many-body quantum system by experimentally demonstrating the Loschmidt echo and accessing OTOCs for local Pauli operators. The quantum system is implemented with a lattice of coupled superconducting qubits~\cite{Kjaergaard2020}, where qubit excitations correspond to particles in the quantum system. OTOCs have been recently measured with a 1d chain of trapped ions~\cite{Gaerttner2017}, in 1d and spin-environment nuclear magnetic resonance systems~\cite{Li2017,Wei2018,Niknam2020}, and with a 2d lattice of superconducting qubits using a purely gate-based approach~\cite{Mi2021}. In contrast, our scheme exploits one of the key strengths of superconducting circuits relative to other qubit platforms, namely the combination of resource-efficient analog quantum simulation steps with high-fidelity single- and two-qubit gates. This digital-analog simulation scheme~\cite{Lamata2018} enables us to engineer Hamiltonians and thereby to implement a reverse time evolution -- the central building block to observe a Loschmidt echo and measure OTOCs. In addition, superconducting circuits provide unique advantages for studying OTOCs and the dynamics of many-body quantum physics in comparison to atomic qubit platforms: they offer high experimental fidelity in simultaneous, site-selective state preparation, control, and readout as well as state and process tomography. Furthermore, superconducting qubit lattices can be readily fabricated with a native 2d connectivity.

Our method is not limited to the single-particle scenario, where information propagation can be inferred by recording the particle distribution within the lattice as a function of time, but enables us to study many-body effects in interacting lattices. Furthermore, we demonstrate that the method extends to disordered lattices, such that it can be used to probe violations of the eigenstate thermalization hypothesis and the breakdown of ergodic behavior~\cite{Palmer1982} in strongly interacting many-body systems. This regime was coined many-body localization~\cite{Huse2014,Nandkishore2015,Roushan2017,Altman2018,Alet2018,Wei2018,Lukin2019,Abanin2019,Chiaro2019} and has so far been predicted and observed for disordered, interacting many-body systems in 1d as an extension of Anderson's original localization phenomenon~\cite{Anderson1958} in non-interacting systems.

\section{Experimental system}

In our experiments, we use a 2d lattice of nine capacitively coupled, superconducting transmon qubits~\cite{Koch2007a}, as schematically depicted in Fig.~\ref{fig:fig1}B. The circuit implements the 2d Bose-Hubbard model~\cite{Yanay2020,Kjaergaard2020}, described in the laboratory frame by
\begin{equation}
\hat H _{\mathrm{BH}}/\hbar =-\sum_{\langle i,j\rangle}J_{ij}\hat a _i^\dagger \hat a _j+ \sum_i\left[\omega_i\hat n _i+\frac{U_i}2\hat n _i(n_i-1)\right]
\label{eq:bosehubbard}
\end{equation}
where $\hat a _i^\dagger$ ($\hat a _i$) is the creation (annihilation) operator for a boson at site $i$, and $\hat n _i=\hat a _i^\dagger \hat a _i$ is the respective particle number operator. The first term describes the hopping interaction between neighboring lattice sites with strength $J_{ij}$, with particle non-conserving terms omitted in a rotating wave approximation ($J_{ij}\ll \omega_i$). The second term represents the on-site energies $\omega_i$, which are given by the transmon transition frequencies. The last term accounts for the anharmonicities $U_i$ of the transmon qubits, representing the energy cost for multiple particles to occupy the same site. A micrograph of the sample used in our experiment is shown in Fig.~\ref{fig:fig1}C. The nearest neighbor coupling rates and qubit anharmonicities have minor deviations from their respectively uniform target values $J/2\pi=\SI{8.1}{MHz}$ and $U/2\pi=-\SI{0.244}{GHz}$. While they are fixed through sample design and remain constant in all our experiments, we can individually tune the on-site energies in a range $\SI{3}{GHz}\lesssim\omega_i/2\pi\lesssim\SI{5.5}{GHz}$ on a timescale much shorter than $1/J$ and the coherence time of our circuit~\cite{suppl}. This degree of control enables us to dynamically switch the time evolution of the system on and off, and allows for the application of single and two-qubit gates. Furthermore, our experimental setup features site-resolved, multiplexed single-shot dispersive qubit readout~\cite{Blais2004,Wallraff2004} with an average qubit state assignment fidelity of $94\%$~\cite{suppl}.

Since $J\ll |U|$ in our circuit, the system always remains in the qubit manifold. This yields the hard-core Bose-Hubbard model, where each lattice site can either be empty or occupied by a single particle~\cite{Yanay2020}. The effective system Hamiltonian in the frame rotating at the common reference frequency $\wq/2\pi=\SI{5.3}{GHz}$ becomes
\begin{equation}
\hat H/\hbar=-\sum_{\langle i,j\rangle}J_{ij}\hat\sigma _i^+ \hat\sigma _j^- +\sum_i\frac{\Delta\omega_i}2\hat\sigma _i^z
\label{eq:hardcoreBH}
\end{equation}
where $\hat\sigma _i^+$ ($\hat\sigma _i^-$) is the raising (lowering) operator for a qubit on site $i$ and $\hat\sigma _i^z$ is the Pauli-$Z$ matrix. We have also defined the rotating frame qubit frequencies $\Delta\omega_i=\omega_i-\wq$. While the on-site interaction term is does not appear in Eq.~\ref{eq:hardcoreBH}, this does not imply the system is non-interacting. The finite on-site interaction term is replaced with a `hard-core' constraint -- lattice sites can be occupied by a single particle at most. The hard-core Bose-Hubbard model is non-integrable in 2d, and therefore signatures of many-body localization are expected even in the absence of a finite on-site interaction term.

\section{Rewinding time\label{sec:timerev}}

The key challenge in our experiments is the realization of both forward $\hat U (t)$ and backward $\hat U (\shortminus t)$ continuous time evolution, a prerequisite for studying time reversibility with a Loschmidt echo and for accessing OTOCs. Forward time evolution is straightforwardly implemented by bringing all qubits on or close to resonance such that the natural time evolution of our quantum system leads to the emulation of $\hat U (t)=\exp[-i\hat Ht]$. 

We achieve the reverse time evolution by constructing a Hamiltonian with inverted sign, a technique that was similarly proposed to experimentally access OTOCs~\cite{Swingle2016} and used in nuclear magnetic resonance experiments~\cite{Li2017,Sanchez2020}. For this, we sandwich a forward time evolution step $\hat U \pr(t)$, under the Hamiltonian $\hat H \pr$, between two sets of single qubit gates $\SZ=\Pi_{i\in\mathrm{black}} \hat\sigma _i ^z$, which are Pauli-$Z$ gates applied to every other qubit in the lattice in both dimensions (i.e.~to all black qubits in a checkerboard pattern~\cite{suppl}). The Hamiltonian $\hat H \pr$ is identical to $\hat H$, but with flipped disorder frequencies, $\Delta\omega_i\rightarrow\shortminus\Delta\omega_i$. In the hard-core Bose-Hubbard model (Eq.~\ref{eq:hardcoreBH}), we find $\SZ\hat \sigma^+_i\hat\sigma^-_j\SZ=-\sigma^{+}_i\hat\sigma^{-}_j$ for any pair of adjacent qubits $\langle i,j \rangle$, and therefore
\begin{equation}
\SZ\hat H \pr\SZ/\hbar=\sum_{\langle i,j\rangle}J_{ij}\hat\sigma _i^+ \hat\sigma _j^- +\sum_i\frac{\shortminus\Delta\omega_i}2\hat\sigma _i^z=-\hat H/\hbar.
\end{equation}
The natural time evolution now yields
\begin{equation}
\SZ\hat U \pr(t)\SZ=\SZ e^{-i\hat H \pr t}\SZ=e^{\shortminus i(\shortminus\hat H)t}=\hat U (\shortminus t),
\end{equation}
realizing the required time-reversed evolution~\cite{suppl}.

The circuit diagram for implementing ${\hat W _j(t)=\hat U (\shortminus t)\hat W _j\hat U (t)}$ is depicted in Fig.~\ref{fig:fig1}D. After a forward time evolution $\hat U (t)$, we apply the local perturbation $\hat W _j$ to lattice site $j$, followed by a backwards time evolution $\hat U (\shortminus t)$. In the context of OTOCs, $\hat W _j$ is called the butterfly operator, reminiscent of the classical butterfly effect in chaos theory, where the state of non-linear systems can be highly sensitive on a small local change~\cite{Lorenz1993}.

\section{Loschmidt echo}

\begin{figure*}
\subfloat{\label{fig:loschsequence}}
\subfloat{\label{fig:loschfidelity}}
\subfloat{\label{fig:loschrho0tom}}
\subfloat{\label{fig:loschrho70tom}}
\subfloat{\label{fig:loschconcurphi}}
\subfloat{\label{fig:loschconcurtau}}
\includegraphics{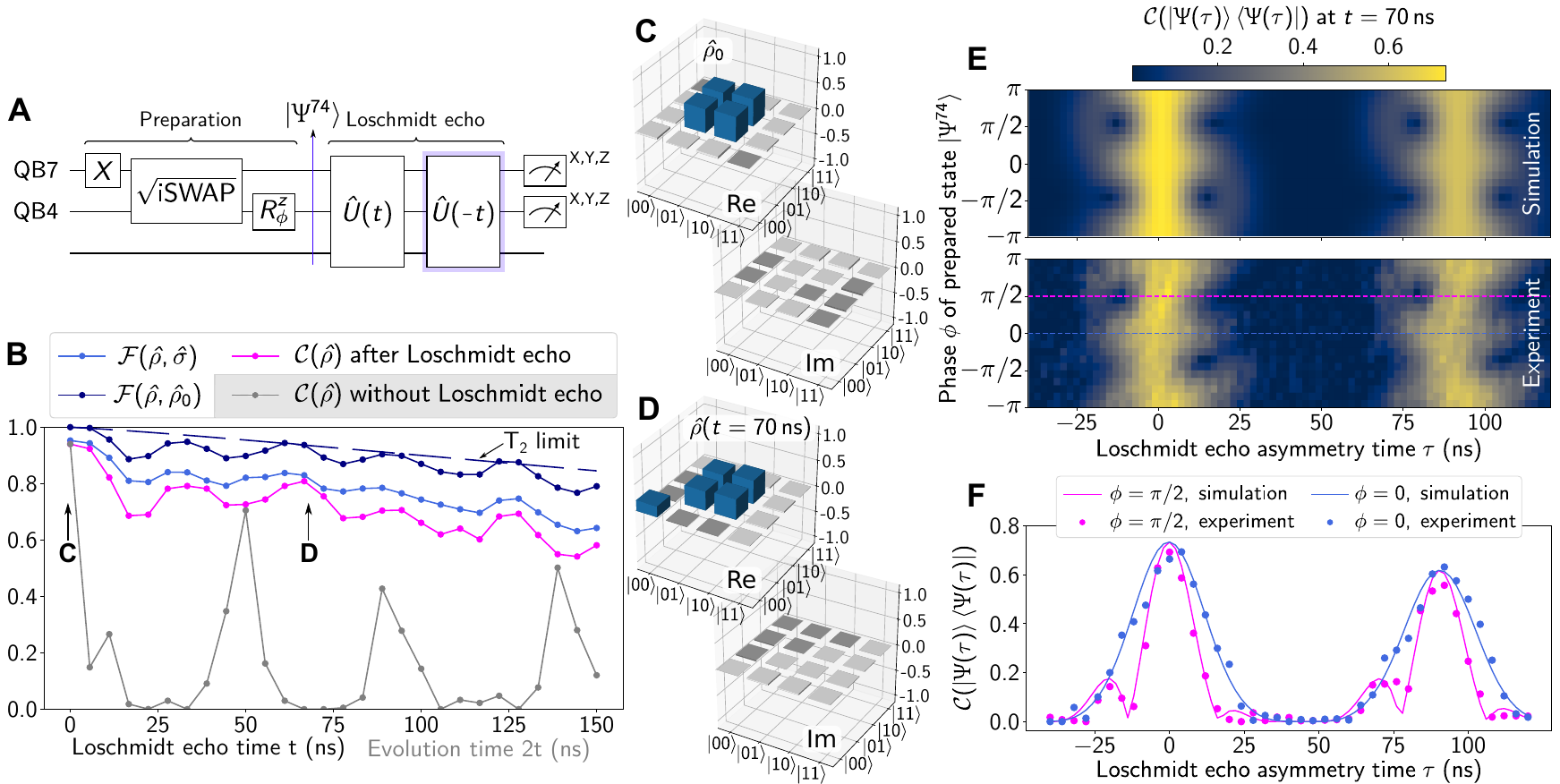}
\caption{
\textbf{Loschmidt echo}
(\textbf{A}) Pulse sequence used in the Loschmidt echo experiment.
(\textbf{B}) Overlap fidelity $\fidelity$ of the extracted density matrix $\hat\rho$ after varying Loschmidt echo times $t$ with the ideal entangled state $\hat\sigma=\Ket{\Psi^{74}}\Bra{\Psi^{74}}$ (bright blue), and with the prepared entangled state $\hat\rho _0$ (dark blue) as measured immediately after state preparation. The coherence limit on $\fidelity(\hat\rho,\hat\rho _0)$ due to qubit dephasing is plotted as a dashed line, based on a net $T_{2,\mathrm{eff}}=\SI{0.88}{\micro s}$~\cite{suppl}. We also show the concurrence of the measured state after a Loschmidt echo (pink) and for comparison without the time-reversal operator in the middle, but with two successive forward time evolution steps $\Ket{\Psi (t)}=\hat U (t)\hat U (t)\Ket{\Psi^{74}}$ (gray).
(\textbf{C}) Reconstructed density matrix of the experimentally prepared state $\hat\rho _0$, $\fidelity(\hat\rho _0,\hat\sigma)=95.3\%$.
(\textbf{D}) Reconstructed density matrix of the measured state $\hat\rho (t=\SI{70}{ns})$, $\fidelity(\hat\rho,\hat\rho _0)=91.5\%$.
(\textbf{E}) Concurrence of the initially entangled subspace after a sequence with the reverse evolution time offset by $\tau$, $\Ket{\Psi(\tau)}=\hat U (\shortminus\SI{70}{ns}-\tau)\hat U (\SI{70}{ns})\Ket{\Psi^{74}}$. We observe an entanglement revival after a Loschmidt echo with $\tau=0$. The additional revival at $\tau=T$ reveals the lattice periodicity.
(\textbf{F}) Line-cuts from the data set in E.
}
\label{fig:loschmidt}
\end{figure*}

\begin{figure*}
\subfloat{\label{fig:otocQRW}}
\subfloat{\label{fig:otocOTOCexp}}
\subfloat{\label{fig:otocOTOCsim}}
\includegraphics{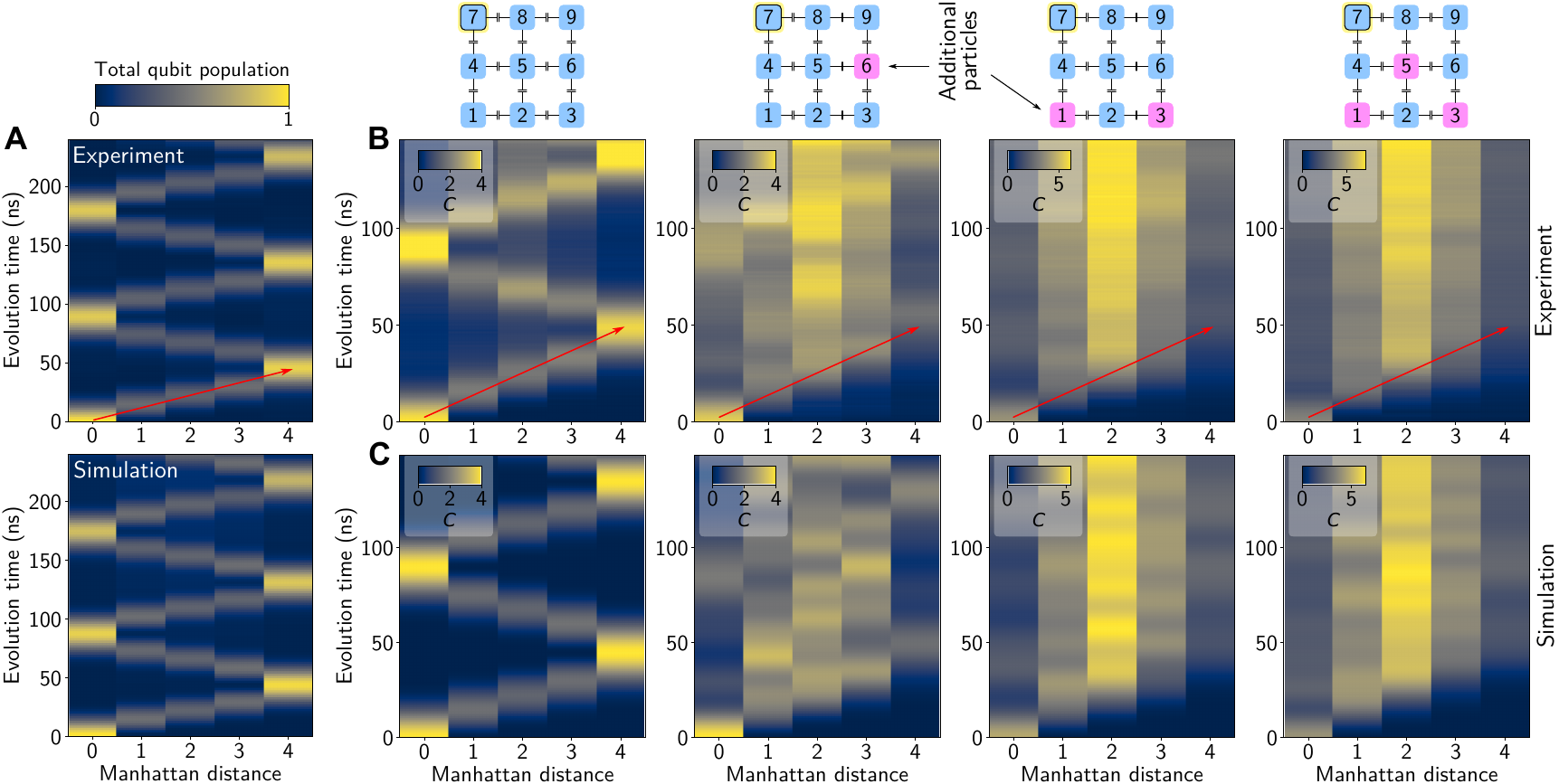}
\caption{
\textbf{Quantum random walk and information propagation via OTOC measurements}
(\textbf{A}) Quantum random walk with a single particle injected at a corner qubit (QB7). We observe a linear light cone and periodic particle propagation.
(\textbf{B}) OTOC measurements for varying number of additional particles (purple) in the lattice, with the linear light cones persisting independent of the number of particles present. We plot the constructed squared commutator $\comm=2+\comm_-+\comm_+$, corrected for dephasing~\cite{suppl}.
(\textbf{C}) Corresponding numerical simulations of $\comm$ based on Eq.~\ref{eq:hardcoreBH} and in the absence of decoherence.
}
\label{fig:otoc}
\end{figure*}

We first investigate the time reversibility of our lattice by performing a quantum Loschmidt echo~\cite{Prosen2003}: the information in a sub-system first disperses throughout the rest of the lattice, which now can be considered as its bath, and then is recovered after reversing the time evolution of the system. We initially prepare the entangled state
\begin{equation}
\Ket{\Psi^{74}}=\frac 1{\sqrt{2}}\left(\Ket{\mathrm{e}}_7\Ket{\mathrm{g}}_4+e^{\I\phi}\Ket{\mathrm{g}}_7\Ket{\mathrm{e}}_4\right)\otimes \prod_{i\ne4,7}\ket{\rm g}_i,
\label{eq:entstate}
\end{equation}
where $\Ket{\mathrm{g}}_{i}$ ($\Ket{\mathrm{e}}_{i}$) denotes the ground state (excited state) of qubit $i$. The two-qubit sub-system consisting of qubits seven and four (yellow frame in Fig.~\ref{fig:fig1}B) is prepared in a Bell state up to a phase $\phi$ that we control during state preparation via a rotation gate $R_\phi^z$, see the pulse sequence in Fig.~\ref{fig:loschmidt}A. We then apply a forward time evolution step $\hat U (t)$, immediately followed by a backward time evolution step $\hat U (\shortminus t)$. The latter is realized according to the construction given in Fig.~\ref{fig:fig1}D, with the final $\SZ$ operation applied as virtual $Z$-gates~\cite{McKay2017} by absorbing them into the subsequent tomography pulses. Finally, we reconstruct the density matrix $\hat \rho$ of the final state after varying Loschmidt echo times via two-qubit tomography in the initially entangled subspace. In Fig.~\ref{fig:loschmidt}B, we show the overlap fidelity
\begin{equation}
\fidelity(\hat\rho,\hat\sigma)=\left[\tr\sqrt{\sqrt{\hat\rho}\hat\sigma\sqrt{\hat\rho}}\right]^2
\end{equation}
between $\hat\rho$ and the density matrix of the ideal entangled state, $\hat\sigma=\Ket{\Psi^{74}}\Bra{\Psi^{74}}$ (Eq.~\ref{eq:entstate}). We also plot the overlap fidelity $\fidelity(\hat\rho,\hat\rho _0)$ with the density matrix of our experimentally prepared state $\hat\rho _0 = \hat \rho(t=0)$, with $\fidelity(\hat\rho _0,\hat\sigma)=95.3\%$ (Fig.~\ref{fig:loschmidt}C). Since the qubits are mutually detuned during state tomography, they acquire single-qubit phase rotations. We therefore find the optimum relative phase in the measured $\hat\rho (t)$ prior to calculating the overlap fidelities~\cite{suppl}. From representative tomography data taken after a Loschmidt echo with $t=\SI{70}{ns}$ (Fig.~\ref{fig:loschmidt}D), we find $\fidelity(\hat\rho,\hat\rho _0)=91.5\%$. The observed degradation in $\fidelity(\hat\rho,\hat\rho _0)$ is in agreement with the coherence limit in our circuit (dashed line in Fig.~\ref{fig:loschmidt}B), which we calculate based on a net dephasing time $T_{2,\mathrm{eff}}=\SI{0.88}{\micro s}$, as extracted from OTOC data in Fig.~\ref{fig:otoc}~\cite{suppl}. In addition, we calculate the concurrence $\mathcal{C}(\hat\rho)$ for the measured states in the two-qubit subspace after the Loschmidt echo. The concurrence is an entanglement metric defined for a mixed state of two qubits and is therefore independent of single-qubit rotations of $\hat\rho$~\cite{Hill1997,suppl}. In Fig.~\ref{fig:loschmidt}B we observe the slow decrease of the concurrence for increasing Loschmidt echo times. In contrast, the concurrence sharply drops to almost zero if we instead apply a sequence yielding two successive forward time evolution steps $\hat U (t)\hat U (t)\Ket{\Psi^{74}}=\hat U (2t)\Ket{\Psi^{74}}$. The periodic peaks in this concurrence trace at integer multiples of $T/2$ reflect the period $T\approx\SI{90}{ns}$ of the quantum random walk in our 2d lattice (see Fig.~\ref{fig:otoc}).

In order to further demonstrate the reversibility of quantum evolution, we observe the revival of entanglement in the two-qubit subspace after a successful Loschmidt echo. In Fig.~\ref{fig:loschmidt}E we vary the backwards evolution time $t+\tau$ after a fixed forward evolution time $t=\SI{70}{ns}$. Entanglement is restored for $\tau=0$ with sub-structure in the line shape dependent on the phase $\phi$ of the prepared state (Eq.~\ref{eq:entstate}). The second revival at $\tau=T$ again reveals the periodicity of the quantum random walk in our lattice. Line cuts for prepared states with $\phi=0,\pi/2$ (Fig.~\ref{fig:loschmidt}F) show that all features in the observed entanglement revival are in good agreement with theory.

\section{Probing information propagation with OTOCs}

In a lattice with only one particle present, the propagation of information corresponds to the propagation of that particle. We measure its quantum random walk (Fig.~\ref{fig:otoc}A) in the absence of lattice disorder ($\Delta\omega_i=0$) after injecting a single particle at a corner of our lattice (qubit seven in Fig.~\ref{fig:fig1}B). Since the propagation in the 2d lattice is along its diagonal symmetry axis (red arrow in Fig.~\ref{fig:fig1}B), we plot the sum of populations in all qubits at a given Manhattan distance from the origin of the random walk. We find a linear light cone, consistent with ballistic particle propagation, and a periodicity of $T\approx\SI{90}{ns}$, dictated by the coupling strength $J$ and the boundary conditions of the lattice. Due to symmetry, our 2d random walk is qualitatively similar to a 1d random walk~\cite{Ma2019,Yan2019} in a chain of length five, however we observe a quadratic increase in the number of traversed lattice sites~\cite{Gong2021,Karamlou21}, a unique signature in 2d. We find excellent agreement with numerical simulations by assuming the same net dephasing time as in Fig.~\ref{fig:loschmidt}B. While the coupling-induced flatness of qubit spectra with respect to any noise variable leads to the protection effect known from small-gap qubits~\cite{Campbell2020}, we observe only a moderate increase of the effective dephasing times in our multi-qubit experiment.

This approach of equating information propagation with particle propagation fails in an interacting lattice with more than one particle present, due to the indistinguishability of bosons. A generic method for probing the propagation of information and other properties of interacting lattices in any dimension is provided by measuring suitable OTOCs~\cite{Swingle2018}. In our experiments, we use
\begin{equation}
\otoc(t) = \langle \hat\sigma _j^z(t) \hat \sigma_i^x \hat\sigma _j^z(t) \hat \sigma_i^x\rangle
\label{eq:ourotoc}
\end{equation}
to probe the information light cone irrespective of the number of particles present. The full information about $\otoc$ is contained in the squared commutator $\comm$, which is Hermitian by construction (Eq.~\ref{eq:otocdef}) and is therefore an experimental observable. Since $\comm$ is itself out-of-time ordered, we construct it according to $\comm=2+\comm_--\comm_+$~\cite{suppl} from two successive measurements yielding $\comm_\pm$ with initial perturbations $(1\pm\hat\sigma _i^x)/\sqrt 2$, respectively, and the butterfly operation $\hat W _j=\hat\sigma _j^z$ applied at time $t$. The pulse sequences are shown in Fig.~\ref{fig:fig1}D, with $\hat V _i=\hat\sigma _i^x$ and $\hat W _j=\hat\sigma _j^z$. The observables $\comm_\pm$ contain time-ordered operator sequences, however they still include the non-unitary operation $(1\pm\hat\sigma _i^x)/\sqrt 2$. We realize this with a unitary rotation gate by restricting the initial state of qubit $i$ to its ground state~\cite{suppl}.

\begin{figure*}
\subfloat{\label{fig:disorderotoc1}}
\subfloat{\label{fig:disorderlightcone}}
\includegraphics{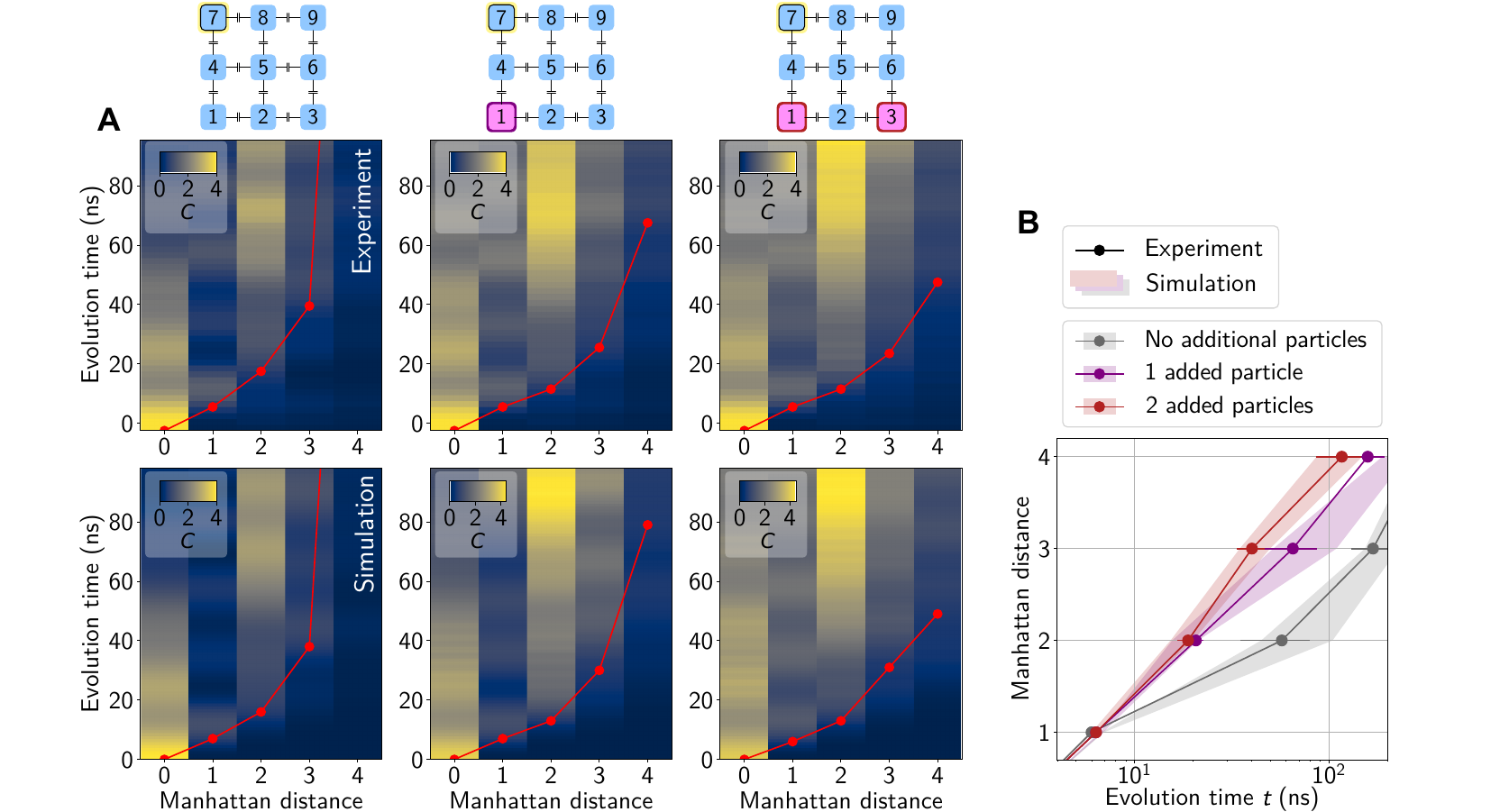}
\caption{
\textbf{OTOC measurements in disordered lattices}
(\textbf{A}) Representative OTOC data in a disordered lattice with $\langle\langle(\Delta\omega _i)^2\rangle\rangle^{1/2}=2.7J$ and a varying number of particles in the lattice. Light cones (red) are extracted by thresholding measured OTOC data~\cite{suppl}. The top row shows experimental data that is corrected for the effects of qubit dephasing~\cite{suppl}, while the bottom row shows numerical simulations of our experimental Hamiltonian. If the threshold is not reached within the simulated $\SI{100}{ns}$, we assign a time of $\SI{300}{ns}$~\cite{suppl}. In the absence of additional particles (left column), we observe that information is localized: the light cone is cut off and information does not reach sites at distance $4$. Localization is slowly overcome in the presence of more particles.
(\textbf{B}) Averaged light cone results for twelve random disorder realizations with identical disorder strength for experiment (dots) and numerical simulation (shaded regions). By adding particles to the lattice, the propagation speed increases. Error bars and the width of the shaded regions show the standard deviation of the mean. By performing numerical simulations for $100$ additional random disorder configurations with $\langle\langle(\Delta\omega _i)^2\rangle\rangle^{1/2}=2.7J$, we confirm that the random realizations used in experiment form a representative set for the observed effect.
}
\label{fig:otocdisorder}
\end{figure*}

In order to probe information propagation in our system, we perform nine pairs of measurements, applying the initial perturbation to qubit $i=7$, but varying the site $j$ which receives the butterfly perturbation. We then sort the measurement results according to their Manhattan distance from $i$ and sum them. We plot OTOC measurements for the degenerate lattice and for different initial states with a varying number of additional particles added to the lattice (Fig.~\ref{fig:otoc}B). We observe that the form of the light cone (red arrows) is independent of the number of particles. Blurring of the propagation features, in the presence of additional particles, is in accordance with numerical simulations of $\comm(t)$ (Fig.~\ref{fig:otoc}C) for the hard-core Bose-Hubbard Hamiltonian in Eq.~\ref{eq:hardcoreBH}.

\section{Information propagation in disordered lattices}

We have so far demonstrated our OTOC protocol for a degenerate lattice, with all qubits at the common reference frequency $\wq$ during time evolution. Next, we apply our method to investigate lattices in the presence of frequency disorder, revealing a change in the speed of information propagation dependent on the number of particles present. For this, we extract OTOC data for twelve random disorder realizations $\{\Delta\omega _i\}$, each with a mean $\langle\langle\Delta\omega _i\rangle\rangle=0$ and a target standard deviation $\langle\langle(\Delta\omega _i)^2\rangle\rangle^{1/2}=2.7J$. We perform OTOC measurements for a varying number of particles in the lattice and for each disorder realization individually, and plot our results for one representative disorder realization in Fig.~\ref{fig:otocdisorder}A. We observe that the presence of disorder hampers the propagation of information: in non-interacting systems, or equivalently when only a single particle is present, information remains localized and never reaches a portion of the lattice at sufficiently strong disorder. This phenomenon, known as Anderson localization~\cite{Anderson1958}, is well understood theoretically, and has been observed in previous studies of random walks in 1d spin chains~\cite{Wei2018}. In our 2d lattice, we find that higher degrees of disorder are necessary to inhibit transport. Interacting systems display a more complicated localization behavior, generally referred to as many-body localization~\cite{Nandkishore2015}. In the presence of more than one particle, we observe that information remains localized at small time scales, but is more likely to eventually reach all parts of the lattice.

In order to quantify the effects of interaction on the spread of information, we extract a `light cone' from each measured OTOC, defined by the time when a threshold value of $0.6$ is first reached at a given Manhattan distance~\cite{suppl} (red dots and lines in Fig.~\ref{fig:otocdisorder}A). Averaging the light cone data from all disorder realizations and for each lattice filling level, respectively, yields a measure for the speed of information propagation in the disordered lattices (Fig.~\ref{fig:otocdisorder}B). We find that this speed of light increases with increasing number of particles, which we construe as a signature of many-body localization in our small 2d lattice: as more particles are added, the increased interactions slowly overcome localization. For the effectively non-interacting lattice (single particle), information propagation eventually stalls, however it approaches a logarithmic time dependence for more particles present (Fig.~\ref{fig:otocdisorder}B).
The observed behavior is characteristic of many-body localized systems~\cite{Huse2014,Nandkishore2015} and is in qualitative agreement with numerical simulations of the 2d hard-core Bose-Hubbard model~\cite{suppl}. Such `interaction-assisted' phenomenon that overcomes localization is not present in 1d~\cite{suppl}, where the hard-core Bose-Hubbard model maps to free fermions~\cite{Jordan1928}.

\section{Conclusion}

We have demonstrated an experimental method to extract OTOCs in strongly interacting many-body systems, which can be generalized to lattices of arbitrary dimension and to various experimental quantum hardware platforms. Our method relies on the application of forward and backward time evolution steps, which we achieve by interleaving blocks of unitary time evolution and single-qubit gates. We have extracted OTOCs that enable us to study quantum information propagation in lattices with a various number of particles present. Our technique further applies to frequency-disordered lattices, which has enabled us to observe an increase in the speed of information propagation with more particles added -- a signature of many-body localization in the 2d hard-core Bose-Hubbard model.

Applying the presented technique to larger lattices has the potential to further our understanding of quantum thermodynamics and black hole dynamics~\cite{Shenker2014,Hayden2007}, and to examine the use of many-body systems for quantum memories~\cite{Nandkishore2015}. In addition, experimentally accessing OTOCs in large quantum circuits may provide a powerful benchmarking tool to study future quantum processors.

\section*{Acknowledgments}

The authors are grateful to P.~M.~Harrington, A.~Di Paolo, S.~Muschinske, and B.~Swingle for insightful discussions, and to M.~Pulido and C.~Watanabe for administrative assistance. 
MK acknowledges support from the Carlsberg Foundation during part of this work. AHK acknowledges support from the NSF Graduate Research Fellowship Program. 
This research was funded in part by the U.S.~Army Research Office Grant W911NF-18-1-0411 and the Assistant Secretary of Defense for Research \& Engineering under Air Force Contract No. FA8721-05-C-0002. Opinions, interpretations, conclusions, and recommendations are those of the authors and are not necessarily endorsed by the United States Government.

\section*{Author Contributions}
JB, AHK, YY, CT, and WDO conceived the experiment. 
YY developed the pulse sequences for rewinding time and extracting OTOCs and generated numerical simulations of the OTOCs.
JB and AHK performed the experiments with theoretical support from YY.
JB, AHK, BK, MK, YS, AV, RW, and SG developed the experiment control tools used in this work.
DK, AM, BN, and JY fabricated the 3x3 qubit array.
TPO, SG, CT, and WDO provided experimental oversight and support. 
All authors contributed to the discussions of the results and to the development of the manuscript.

\bibliography{otoc}

\end{document}


\title{Supplementary material for ``Probing quantum information propagation with out-of-time-ordered correlators''}

\def\RLEaffil{Research Laboratory of Electronics, Massachusetts Institute of Technology, Cambridge, MA 02139, USA}
\def\LLaffil{MIT Lincoln Laboratory, Lexington, MA 02421, USA}
\def\Physaffil{Department of Physics, Massachusetts Institute of Technology, Cambridge, MA 02139, USA}
\def\EECSaffil{Department of Electrical Engineering and Computer Science, Massachusetts Institute of Technology, Cambridge, MA 02139, USA}
\def\Maryaffil{Laboratory for Physical Sciences, 8050 Greenmead Dr., College Park, MD 20740, USA}

\author{Jochen~Braum\"uller}
\email{jbraum@mit.edu}
\affiliation{\RLEaffil}
\author{Amir~H.~Karamlou}
\affiliation{\RLEaffil}
\affiliation{\EECSaffil}
\author{Yariv~Yanay}
\affiliation{\Maryaffil}
\author{Bharath~Kannan}
\affiliation{\RLEaffil}
\affiliation{\EECSaffil}
\author{David~Kim}
\affiliation{\LLaffil}
\author{Morten~Kjaergaard}
\affiliation{\RLEaffil}
\author{Alexander~Melville}
\author{Bethany~M.~Niedzielski}
\affiliation{\LLaffil}
\author{Youngkyu~Sung}
\affiliation{\RLEaffil}
\affiliation{\EECSaffil}
\author{Antti~Veps\"al\"ainen}
\author{Roni~Winik}
\affiliation{\RLEaffil}
\author{Jonilyn~L.~Yoder}
\affiliation{\LLaffil}
\author{Terry~P.~Orlando}
\affiliation{\RLEaffil}
\affiliation{\EECSaffil}
\author{Simon~Gustavsson}
\affiliation{\RLEaffil}
\author{Charles~Tahan}
\affiliation{\Maryaffil}
\author{William~D.~Oliver}
\affiliation{\RLEaffil}
\affiliation{\EECSaffil}
\affiliation{\LLaffil}
\affiliation{\Physaffil}

\date{\today}

\maketitle

\tableofcontents

\section{Sample and experimental setup}

The sample used in our experiments is a 2d lattice of capacitively coupled transmon qubits~\cite{Koch2007a} with floating capacitor pads. The capacitor pads each have two coupling fingers attached, facilitating a nearest-neighbor qubit-qubit interaction with average strength $J/2\pi=(8.1\pm 0.2)\,\mathrm{MHz}$, measured at qubit frequencies of $\SI{5.5}{GHz}$. As shown in Fig.~1C, the transmon capacitor pads share a common symmetry axis spanned from the top left to the bottom right corner of the chip. This ensures that the coupling phases add up to zero when traversing a closed loop of qubits, leading to vanishing effective gauge fields. The coupling strength between qubits one and four (see numbering convention in Fig.~1B) deviates from the mean by $8\%$ due to the crossing of the central qubit's readout resonator.

The next-nearest neighbor interaction in our qubit lattice is a result of both the direct capacitance between nonadjacent qubits, and an effective contribution due to the nearest neighbor couplings in the circuit's capacitance matrix. Due to the floating transmon geometry and the equal rotation of the capacitor pads, the effective contribution is stronger along the top-left to bottom-right corner of our lattice, since next-nearest neighbor qubits couple via the same capacitor pads of the intermediate qubit, resulting in a anisotropy of the next-nearest neighbor coupling. By calculating the complete capacitance matrix of our circuit, we find the largest non-nearest neighbor couplings to be $\sim J/75$.

Apart from lithographically defined cross-over bridges used for the hopping of signal lines over other circuit elements and to connect the ground planes, the circuit is realized in a single-chip planar design. Qubit microwave manipulation (XY control) and multiplexed dispersive state readout is performed through individual capacitively coupled coplanar resonators, which are in turn coupled to a common single-port Purcell filter. We use $\lambda/4$ readout resonators for all qubits along the edge of the lattice to minimize their footprint on the sample. The central qubit is read out with an `open-terminated' $\lambda/2$ resonator that crosses over one of the corner qubits. In order to minimize their parasitic coupling, the crossing is located in the middle of the $\lambda/2$ resonator at its voltage node. The Purcell filter is designed to have a large bandwidth of $\SI{0.54}{GHz}$ at a resonance frequency of $\SI{7.3}{GHz}$, such that the readout resonators can be distributed over a $\sim\SI{400}{MHz}$ band. This non-aggressive design of the Purcell filter enables us to manipulate qubits through the readout port while limiting Purcell decay to a rate $\lesssim1/(\SI{500}{\micro s})$. With effective resonator linewidths in the range of $(1-3)\,\mathrm{MHz}$ at the readout frequency, we achieve an average qubit state assignment fidelity of $94\%$ at an integration time of $\SI{800}{ns}$.

Each qubit is coupled to a flux bias line used for both static flux biasing and pulsed flux control (`fast flux control' or Z control) with a maximum bandwidth of $\sim\SI{400}{MHz}$. Purcell loss through the flux bias lines limits the $T_1$ times to $\approx\SI{35}{\micro s}$ due to an asymmetry of the flux lines with respect to the transmon capacitor pads. This main loss channel can be remedied in future sample generations by improving the design of the flux line. The mutual inductance between the qubits' SQUID loops and their respective flux bias line is $\SI{0.5}{pH}$. Flux cross-talk calibrations for both dc flux control and fast flux control are provided in Sec.~\ref{sec:dcxtalk} and Sec.~\ref{sec:fastxtalk}.

Detailed sample parameters for individually isolated qubits are summarized in Tab.~\ref{tab:device-parameters}.

\begin{table}[b]
\centering
\caption{\textbf{Sample parameters}. We show the maximum transmon transition frequencies $\omega_{\mathrm{q}}^{\rm max}$ at the upper flux insensitive point, the qubit anharmonicities $U$ (measured at $\omega_{\mathrm{q}}^{\rm max}$), the readout resonator frequencies $\omega_{\mathrm{r}}$, the probabilities $f_{ij}$ of measuring the qubit in state $i$ after preparing it in state $j$, the readout assignment fidelity $(f_{\mathrm{gg}}+f_{\mathrm{ee}})/2$, and the average measured $T_1^{\rm avg}$ at qubit frequencies around the bias point used in experiment.}
\vspace{8pt}
\label{tab:device-parameters}
{\renewcommand{\arraystretch}{1.7}   
\begin{tabular}{ p{4.5cm}  p{1cm} p{1cm}  p{1cm}  p{1cm} p{1cm} p{1cm}  p{1cm}  p{1cm} p{1cm}   }
\toprule
   &    Q1  &          Q2  &          Q3  &          Q4  &          Q5  &          Q6  &          Q7  &          Q8  &          Q9  \\
\hline
$\omega_{\mathrm{q}}^{\rm max}/2\pi$ (GHz) &  5.712 & 5.771 & 5.788 & 5.707 & 5.822 & 5.626 & 5.528 & 5.673 & 5.722  \\

$U/2\pi$ (MHz) &  -238.7 & -239.6 & -238.2 & -240.0 & -233.6 & -273.8 & -244.8 & -241.4 & -241.1 \\

$\omega_{\mathrm{r}}/2\pi$ (GHz) & 7.221 & 7.170 & 7.121 & 7.304 & 6.942 & 7.324 & 7.208 & 7.28 & 7.155    \\

$f_{\mathrm{gg}}$ & 0.97 & 0.98 & 0.98 & 0.98 & 0.98 & 0.98 & 0.98 & 0.94 & 0.98 \\

$f_{\mathrm{ee}}$ & 0.93 & 0.93 & 0.89 & 0.94 & 0.93 & 0.93 & 0.92 & 0.89 & 0.94 \\

Readout assignment fidelity & 0.94 & 0.94 & 0.93 & 0.96 & 0.95 & 0.96 & 0.95 & 0.91 & 0.96  \\

$T_1^{\rm avg}$ ($\SI{}{\micro s}$)&  12.4 & 14.7 & 13.6 & 11.6 & 13.5 & 11.0 & 8.5 & 15.0 & 12.9  \\

\hline
\hline
\end{tabular}
}
\end{table}

\begin{figure*}
\includegraphics{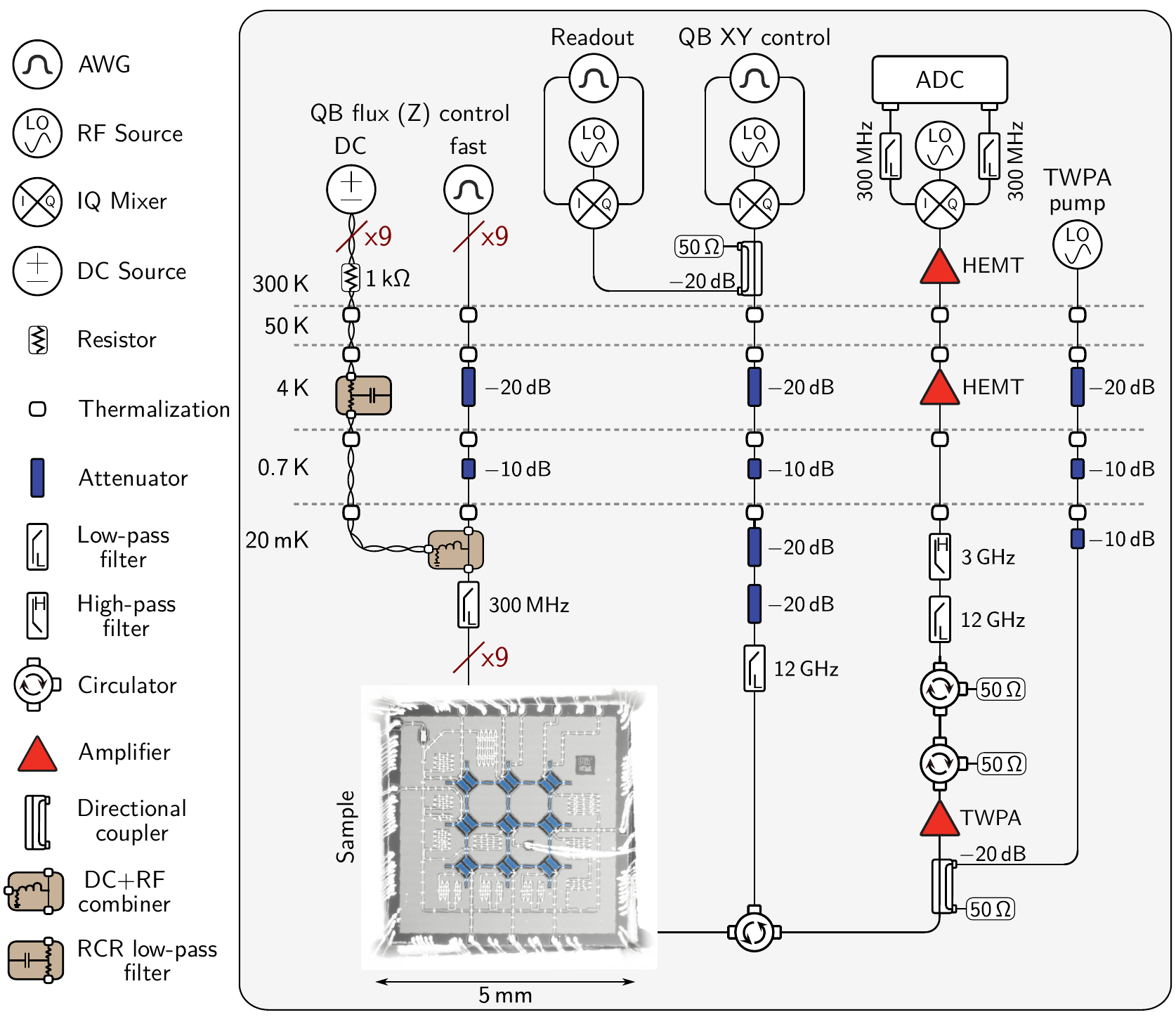}
\caption{
\textbf{Measurement setup}
}
\label{fig:setup}
\end{figure*}

The samples are fabricated on a silicon substrate by dry etching an MBE grown, \SI{250}{nm} thick aluminum film in an optical lithography process, forming all larger circuit elements such as the qubit capacitor pads, and the readout and control circuitry. The qubit SQUID loops are fabricated with an electron beam lithography process and a double angle shadow evaporation technique~\cite{Dolan1977} to form the Josephson junctions. We use a $\SI{5}{\micro m}$ wide wire width for the SQUIDs, which was found to minimize flux noise from local magnetic spin defects on the surfaces and interfaces of the SQUID~\cite{Braumueller2020}.

Our experimental setup is shown in Fig.~\ref{fig:setup}. The sample is cooled down to approximately $\SI{20}{mK}$ in a dilution refrigerator. Our readout setup follows the established heterodyne mixing scheme. The reflected microwave signal is amplified with a travelling wave parametric amplifier (TWPA), which is crucial for our experiment due to its large bandwidth and high saturation level.

We perform qubit control with a heterodyne mixing scheme~\cite{Krantz2019} in order to simultaneously address individual qubits by superimposing different frequency pulses. We calibrate the I and Q quadrature offsets, the gain imbalance, and the skew of our IQ mixer by minimizing both, the local oscillator (LO) leakage and the signal level of the unwanted sideband.

The static flux bias components and the fast flux qubit Z control signals travel to the mixing chamber of our cryostat in twisted-pair dc wires and coaxial cables, respectively. For each qubit, the two signal components are combined into the respective flux bias line by using an rf-choke, consisting of an inductor in the dc input and a $\SI{50}{\ohm}$ shunt capacitor to ground, but no capacitor.

Since the Z control and XY control signals travel through coaxial lines of different length and differing attenuation, the difference in travel time must be compensated by appropriate delays in signal generation. We calibrate the Z-delays for each qubit by applying a short flux pulse with a $\pi$-pulse of the same length and calibrated for the target frequency of the qubit during the flux pulse. When the two pulses are aligned, the population of the qubit is maximal.

\section{Flux cross-talk calibrations}

A central part of our tune-up protocol is the calibration of both the static flux cross-talk and the fast flux cross-talk in our device. We calibrate static flux cross-talk in a three-step process: we first find the static cross-talk matrix through qubit spectroscopy, then we use a learning-based optimization procedure of the cross-talk matrix based on a random sampling of qubit frequency settings, and finally fine-tune the static set qubit frequencies via the measured spectrum of the slightly degenerate, coupled qubits.

Fast flux cross-talk differs substantially from static flux cross-talk. We find the fast flux cross-talk matrix by first using a set of Ramsey measurements and subsequently applying the same optimization procedure.

\begin{figure}
\includegraphics{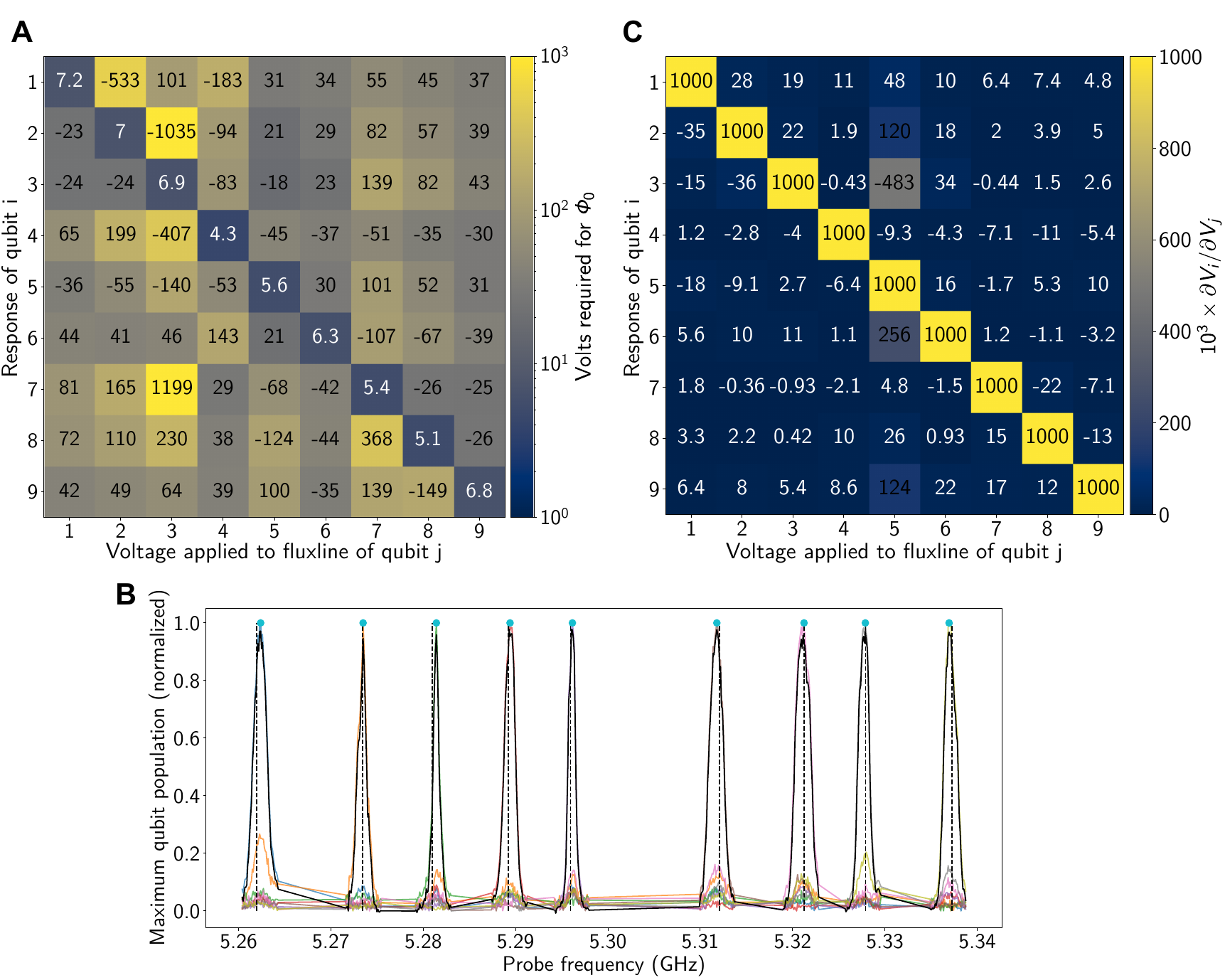}
\caption{
\textbf{Flux cross-talk calibration}
(\textbf{A}) Cross-talk matrix for static flux control after using the learning-based optimization technique. Numbers represent the voltage $V^0_{i,j}$ required to tune qubit $i$ on the vertical axis by one full period (corresponding to applying one flux quantum $\Phi_0$) when applying a voltage to qubit $j$ (horizontal axis).
(\textbf{B}) Fine-tuning of the qubit frequencies via spectroscopy. The nine qubits are detuned by $\SI{8}{MHz}$ from each other, which breaks the degeneracies such that they appear at individual frequencies. The dashed lines indicate the calculated dressed eigenenergies of the single-photon manifold of the Bose-Hubbard Hamiltonian, and the blue points show the measured frequencies in spectroscopy. After optimization, we achieve a frequency standard deviation of about $\SI{0.2}{MHz}$ in the dressed frame.
(\textbf{C}) Cross-talk matrix for fast flux control after optimization. Numbers represent the voltage response $\partial V_i/\partial V_j$ of qubit $i$ in response to a voltage pulse applied to qubit $j$, multiplied by $10^3$.
}
\label{fig:dc_xtalk_matrix}
\end{figure}

\subsection{Static cross-talk matrix}\label{sec:dcxtalk}

For our experiments, we use flux-tunable transmon qubits each with a local flux line for independent flux control. We find an initial static flux cross-talk matrix via qubit spectroscopy of the isolated qubits. The frequency spectrum of transmon $i$ in response to voltage biases applied to the flux lines on the chip is given by~\cite{Koch2007}
\begin{equation}
    \omega_{i} = (\omega_{i}^{\rm{max}}+E_{\mathrm{C},i}) \left[d_i^2 + (1-d_i^2)\: \cos^2 \left( \sum_j\pi\frac{V_j}{V^0_{i,j}}-\phi^0_i \right)\right]^{1/4}-E_{\mathrm{C},i}.
\label{eq:qubitspectra}
\end{equation}
Fit parameters are the maximum frequency $\omega_{i}^{\rm{max}}$ at the flux insensitive point, the asymmetry parameter
\begin{equation}
d_i=[(\omega_i^{\rm{min}}+E_{\mathrm{C},i})/(\omega_i^{\rm{max}}+E_{\mathrm{C},i})]^2    
\end{equation}
of the transmon's dc-SQUID, a phase offset $\phi^0_i$, the transmon charging energy $E_{\mathrm{C},i}$, and the voltage $V^0_{i,j}$ applied to the flux line of qubit $j$ which is required to tune qubit $i$ by a flux quantum $\Phi_0$. These voltages can be translated into currents with our constant bias resistance of $\SI{1.1}{k\ohm}$. The measured static flux cross-talk matrix is shown in Fig.~\ref{fig:dc_xtalk_matrix}A. The diagonal matrix elements correspond to the designed coupling between each qubit and its local flux line, respectively, whereas the off-diagonal elements represent the cross-talk between qubits and flux lines attached to other qubits. We first measure the qubit spectra for $i=j$ and subsequently fix the extracted fit parameters $\omega_i^{\mathrm{max}}$, $d_i$, and $E_{\mathrm{C},i}$ in the fits of the off-diagonal entries, in order to increase the robustness of the fits. The extracted phase offsets vary significantly between different thermal cycles of the sample and average around $\langle\langle|\phi_i^0|\rangle\rangle=0.034$, corresponding to a flux offset of $\approx 0.01\Phi_0$.

Using the fitted qubit spectra and the static flux cross-talk matrix, we can calculate the necessary set voltages for any set of target frequencies by first inverting Eq.~\ref{eq:qubitspectra} to find the diagonal target voltages and then using the inverse of the cross-talk matrix to calculate the necessary voltages to compensate for the cross-talk.

\subsection{Learning-based cross-talk matrix optimization}

The limited voltage range we apply to each flux line results in uncertainty especially in large entries of the cross-talk matrix (having little cross-talk). We improve the previously found cross-talk matrix by using a learning-based gradient descent optimization technique.

We start by initializing our qubits to a random set of target frequencies with a minimum mutual frequency distance of $\sim 2.5J$ using the initial cross-talk matrix obtained from qubit spectroscopy. This yields a set of applied voltages $\vec{V}$ and a set of measured qubit frequencies $\vec\omega _{\mathrm{q}}$. Each qubit in the lattice experiences an ac-Stark shift caused by the interaction with neighboring qubits. We calculate the uncoupled qubit frequencies from the measured dressed frequencies by using the full Bose-Hubbard Hamiltonian (Eq.~2). Using the qubit spectra, we convert these frequencies into a set of fluxes $\vec\Phi$ experienced by each SQUID loop in the system. By repeating this protocol for $N\approx 200$ iterations we obtain a set $\{\vec V _i,\vec\Phi _i\}_{i=1:N}$ of input voltages and the corresponding measured fluxes experienced by the qubits.  

We use this data set $\{\vec V _i,\vec\Phi _i\}_{i=1:N}$ to train our initial cross-talk matrix $\mathbf{M}$, where $\mathbf{M}_{k,l}=1/V^0_{k,l}$. With the offset phases $\vec\varphi$, where $\vec\varphi _k=\phi_k^0$, the sum of the squared differences between the measured fluxes $\{\vec{\Phi}_i\}$ and the value predicted by $\mathbf{M}$, $\vec\varphi$, and the set voltages $\{\vec V _i\}$ can be used as the cost function for a minimization problem, 
\begin{equation}
    c(\mathbf{M},\vec\varphi)=\frac{1}{N} \sum_i^N ||\Vec{\Phi}_i - (\mathbf{M}\vec V _i + \vec\varphi)||^2
\end{equation}
Since the system is described by a linear process with a well-defined convex cost landscape, we minimize $c(\mathbf{M},\vec\varphi)$ with a gradient descent-based optimizer.

\subsection{Spectroscopy-based fine-tuning of qubit frequencies}

In order to evaluate the accuracy of the optimized static cross-talk matrix and in order to fine-tune the qubit frequencies, we perform lattice spectroscopy of the almost fully hybridized lattice. Since the 2d resonant tight binding model has degenerate eigenenergies, we mutually detune the qubits by $\approx\SI{8}{MHz}$ in order to lift the degeneracy and be able to resolve individual peaks. We apply a weak, $\SI{40}{\micro s}$ long microwave pulse of varying probe frequency to the common control line that simultaneously excites all nine qubits. We then far detune the qubits from each other and measure their states. As an example, we show the normalized populations of the summed measurements in Fig.~\ref{fig:dc_xtalk_matrix}B. By comparing the measured eigenenergies (blue dots) with the exact eigenenergies of the single-particle manifold of the Bose-Hubbard Hamiltonian -- which takes the form of a tight-binding model -- we add small frequency offsets that optimize the lattice spectrum. Using this method, we achieve a frequency standard deviation of about $\SI{0.2}{MHz}$ in the dressed frame.

\subsection{Fast flux cross-talk calibration}\label{sec:fastxtalk}

In order to calibrate the fast flux cross-talk matrix, we first extract the qubit spectra for biasing the qubits with short flux pulses. We then individually bias the qubits close to the inflection point of their spectra, where the spectrum is approximately linear and the flux sensitivity is maximal. We use Ramsey measurements to probe the frequency shift of qubit $i$ while we apply a fast flux pulse of voltage $V_j$ to the local flux line of qubit $j$. Using the fast flux transmon spectrum, we calculate the corresponding voltage $V_i$, yielding $\partial V_i/\partial V_j$ for all qubit pairs, see the complete cross-talk matrix in Fig.~\ref{fig:dc_xtalk_matrix}C. We further improve the fast flux cross-talk matrix with the same learning-based optimization technique described in the previous section.

\section{Transient calibration}

Fast flux pulses experience distortions as they travel to the sample through various electric components in the cryostat. By using the qubit as a sensor, we can characterize the distortion in the step response of the flux control line using Ramsey-type measurements. The step response for the signal generated by the AWG, $V_{\rm{AWG}}(t)$, the signal reaching the qubit, $V_{\rm{qubit}}(t)$, and the time-dependent distortion, $n(t)$, are related by
\begin{equation}
    V_{\rm{qubit}}(t) = V_{\rm{AWG}}(t) \: \big(1+n(t) \big),
\end{equation}

To characterize $n(t)$, the target qubit is biased to its sweet-spot using static flux, and excited with a $R^x(\pi/2)$ pulse followed by a $R^z$ rotation pulse for time $t$. The dynamic frequency change of the qubit as a response to the fast $R^z$ pulse can be extracted by measuring both $\langle X \rangle $ and $\langle Y \rangle $, denoting the expectation values of the qubit state projected along the $x$-axis and $y$-axis of the Bloch sphere. The phase accumulated by the qubit during the $Z$ pulse can be described by $\phi (t)=\int_0^t \Delta(t^\prime) \rm{d}t^\prime$, where $\Delta(t)=\Delta_0 + \delta(t)$ is the frequency detuning, consisting of the target detuning $\Delta_0$ and and the added frequency distortion $\delta(t)$. The $\langle X \rangle$ and $\langle Y \rangle$ measurements are used to obtain $\cos\phi(t)$ and $\sin\phi(t)$, respectively, which we use for constructing the phasor $e^{i\phi(t)}$. We can then extract the frequency distortion as $\delta(t)=\frac{\mathrm{d}}{\mathrm{d}t} \arg(e^{i\phi(t)})$. By using the qubit spectrum, we map $\delta(t)$ to the distortion in the signal reaching the qubit $n(t)$. In order to obtain an analytical expression, we fit the extracted $n(t)$ with a sum of typically three exponential damping terms with time constants $\tau_i$ and settling amplitudes $A_i$:
\begin{equation}
    n(t)=\sum_i A_i e^{-t/\tau_i}
\end{equation}
In our experiments, we use the analytical expression of $n(t)$ to pre-distort the output signal of the AWG in order to compensate for the system transients.

\section{Constructing OTOCs from Hermitian variables\label{sec:OTOCconstruction}}

We elaborate here on the experimental process used to access the OTOC defined in Eq.~1, which in our experiment, for $\hat V=\hat\sigma _i^x$ and $\hat W=\hat\sigma _j^z$, takes the form
\begin{equation}
    \otoc(t) = \Bra{\psi} \hat \sigma^{z}_{j}(t) \hat \sigma^{x}_{i} \hat \sigma^{z}_{j}(t) \hat \sigma^{x}_{i}\Ket{\psi}.
    \label{eq:otocAppendix}
\end{equation}

Contrast the form of Eq.~\ref{eq:otocAppendix} with a typical experimental procedure: here, an initial quantum state $\Ket{\psi}$ is prepared, then a series of unitary operators $U_{1},U_{2},\dotsc$ are applied with some time intervals $t_{1},t_{2},\dotsc$ between them, and then finally some Hermitian operator $\hat M$ is measured, yielding, e.g.
\begin{equation}
    \Obs = \bra\psi\hat\Obs\ket\psi=\bra{\psi}\hat U_{1}^{\dagger}(t_{1})\hat U_{2}^{\dagger}(t_{1}+t_{2})\hat M(t_{1}+t_{2}+t_{3})\hat U_{2}(t_{1}+t_{2})\hat U_{1}(t_{1})\Ket{\psi}.
\label{eq:obs}
\end{equation}

The forms of $\Obs$ and $\otoc$ differ in two major ways. First, the operators in $\hat\Obs$ are evaluated at monotonically increasing time values as we move inwards, while, as discussed in the main text, the operators in $\otoc$ are out-of-time ordered. This is resolved by pulse sequences combining forward and backward time evolutions. Second, the complete operator $\hat\Obs$ is itself Hermitian, with each unitary applied on both sides of the measurement operator $\hat M$. In contrast, the operator string in $\otoc$ is non-Hermitian. 
    
We note that the quantity  we seek to measure, $\comm$, \emph{is} Hermitian, as defined in Eq.~1,
\begin{equation}
    \comm=2-\Bra{\psi} \hat \sigma^{z}_{j}(t) \hat \sigma^{x}_{i} \hat \sigma^{z}_{j}(t) \hat \sigma^{x}_{i} + \hat \sigma^{x}_{i} \hat \sigma^{z}_{j}(t) \hat \sigma^{x}_{i} \hat \sigma^{z}_{j}(t)\Ket{\psi}.
\end{equation}
However, this sum cannot be directly converted into a set of gates without fully diagonalizing the Hamiltonian~\cite{Mitarai2019}.

\subsection{Linking $\otoc$ to measurement observables}

In order to make the squared commutator $\comm$ accessible in experiment, we use a mathematical equivalence. We expand
\begin{equation*}\begin{split}
\Bra{\psi}\hat \sigma^{z}_{j}(t) \hat \sigma^{x}_{i} \hat \sigma^{z}_{j}(t) \hat \sigma^{x}_{i}\Ket{\psi} & =
    \Bra{\psi}\hat \sigma^{z}_{j}(t) \hat \sigma^{x}_{i} \hat \sigma^{z}_{j}(t) \frac{1+\hat \sigma^{x}_{i}}{2}\Ket{\psi} - \Bra{\psi}\hat \sigma^{z}_{j}(t) \hat \sigma^{x}_{i} \hat \sigma^{z}_{j}(t) \frac{1-\hat \sigma^{x}_{i}}{2}\Ket{\psi}
    \\ & =\Bra{\psi}\frac{1+\hat \sigma^{x}_{i}}{2}\hat \sigma^{z}_{j}(t) \hat \sigma^{x}_{i} \hat \sigma^{z}_{j}(t) \frac{1+\hat \sigma^{x}_{i}}{2}\Ket{\psi} - 
    \Bra{\psi}\frac{1-\hat \sigma^{x}_{i}}{2}\hat \sigma^{z}_{j}(t) \hat \sigma^{x}_{i} \hat \sigma^{z}_{j}(t) \frac{1-\hat \sigma^{x}_{i}}{2}\Ket{\psi}
    \\ & \quad + \Bra{\psi}\frac{1-\hat \sigma^{x}_{i}}{2}\hat \sigma^{z}_{j}(t) \hat \sigma^{x}_{i} \hat \sigma^{z}_{j}(t) \frac{1+\hat \sigma^{x}_{i}}{2}\Ket{\psi} - 
    \Bra{\psi}\frac{1+\hat \sigma^{x}_{i}}{2}\hat \sigma^{z}_{j}(t) \hat \sigma^{x}_{i} \hat \sigma^{z}_{j}(t) \frac{1-\hat \sigma^{x}_{i}}{2}\Ket{\psi},
\\ \Bra{\psi}\hat \sigma^{x}_{i}\hat \sigma^{z}_{j}(t) \hat \sigma^{x}_{i} \hat \sigma^{z}_{j}(t)\Ket{\psi} & =
    \Bra{\psi}\frac{1+\hat \sigma^{x}_{i}}{2}\hat \sigma^{z}_{j}(t) \hat \sigma^{x}_{i} \hat \sigma^{z}_{j}(t) \frac{1+\hat \sigma^{x}_{i}}{2}\Ket{\psi} - 
    \Bra{\psi}\frac{1-\hat \sigma^{x}_{i}}{2}\hat \sigma^{z}_{j}(t) \hat \sigma^{x}_{i} \hat \sigma^{z}_{j}(t) \frac{1-\hat \sigma^{x}_{i}}{2}\Ket{\psi}
    \\ & \quad + \Bra{\psi}\frac{1+\hat \sigma^{x}_{i}}{2}\hat \sigma^{z}_{j}(t) \hat \sigma^{x}_{i} \hat \sigma^{z}_{j}(t) \frac{1-\hat \sigma^{x}_{i}}{2}\Ket{\psi} - 
    \Bra{\psi}\frac{1-\hat \sigma^{x}_{i}}{2}\hat \sigma^{z}_{j}(t) \hat \sigma^{x}_{i} \hat \sigma^{z}_{j}(t) \frac{1+\hat \sigma^{x}_{i}}{2}\Ket{\psi},
\end{split}\end{equation*}
\begin{equation*}\begin{gathered}
\Rightarrow \comm=2 - \Bra{\psi}\hat \sigma^{z}_{j}(t) \hat \sigma^{x}_{i} \hat \sigma^{z}_{j}(t) \hat \sigma^{x}_{i}\Ket{\psi} - \Bra{\psi}\hat \sigma^{x}_{i}\hat \sigma^{z}_{j}(t) \hat \sigma^{x}_{i} \hat \sigma^{z}_{j}(t)\Ket{\psi} = 
    \\2 - 2\left[ \Bra{\psi}\frac{1+\hat \sigma^{x}_{i}}{2}\hat \sigma^{z}_{j}(t) \hat \sigma^{x}_{i} \hat \sigma^{z}_{j}(t) \frac{1+\hat \sigma^{x}_{i}}{2}\Ket{\psi} - 
    \Bra{\psi}\frac{1-\hat \sigma^{x}_{i}}{2}\hat \sigma^{z}_{j}(t) \hat \sigma^{x}_{i} \hat \sigma^{z}_{j}(t) \frac{1-\hat \sigma^{x}_{i}}{2}\Ket{\psi}\right].
\end{gathered}\end{equation*}

It follows immediately that
\begin{equation}
    \comm = 2 + \comm_{-} - \comm_{+}
    \label{eq:CeqCmCp}
\end{equation}
where
\begin{equation}
    \comm_{\pm} \equiv \Bra{\psi}\frac{1 \pm \hat \sigma^{x}_{i}}{\sqrt{2}} \hat \sigma^{z}_{j}(t) \hat \sigma^{x}_{i} \hat \sigma^{z}_{j}(t) \frac{1 \pm \hat \sigma^{x}_{i}}{\sqrt{2}}\Ket{\psi}.
    \label{eq:Cpmdef}
\end{equation}
These new quantities $\comm_\pm$ are both Hermitian and have the symmetric form of $\Obs$ (Eq.~\ref{eq:obs}), making them experimental observables.

\subsection{Realizing $1\pm \hat \sigma^{x}$ with a restricted initial state}

One experimental obstacle which remains is the realization of $(1\pm \hat \sigma^{x}_{i})/\sqrt{2}$; this is by definition a non-unitary operator, and cannot be realized with a gate sequence within the qubit manifold. We achieve the effect of $(1\pm\hat\sigma ^x)/\sqrt{2}$ by restricting the qubit on site $i$ to its ground state at the beginning of the experiment, using
\begin{equation}
    \frac{1 \pm \hat \sigma ^{x}}{\sqrt{2}}\Ket{\rm g} = \frac{\Ket{\rm g} \pm \Ket{\rm e}}{\sqrt{2}} = \hat R^y\left(\pm \frac{\pi}{2}\right)\Ket{\rm g},
\end{equation}
where $\hat R^y$ denotes a rotation around the $y$-axis of the qubit Bloch sphere. Thus, $\comm_\pm$ is realized via the following sequence:

\begin{tabular}{p{8pt}p{0.5\textwidth}@{\hspace{12pt}}p{0.4\textwidth}}
\\ 1. & Prepare the initial state, with $i$ in the ground state and the rest of the system in some arbitrary state $\Ket{\overline{\psi}}_{\bar i}$,& 
    $\Ket{\psi} = \Ket{\rm g}_{i}\otimes \Ket{\overline{\psi}}_{\bar i}$
\\\\ 2. & Apply the rotation $R_i^y(\pm\frac{\pi}{2})$ to qubit $i$,
    & $\Ket{\psi_{2}} = \frac{\Ket{\rm g}_{i}\pm\Ket{\rm e}_{i}}{\sqrt{2}}\otimes\Ket{\overline{\psi}}_{\bar i}= \frac{1 \pm \hat \sigma^{x}_{i}}{\sqrt{2}}\Ket{\psi}$
\\\\ 3. & Apply a forward time evolution, a $Z$-gate to qubit $j$ as the butterfly operator, and a reverse time evolution, realizing $\hat\sigma _j^z(t)$,
    & $\Ket{\psi_{3}} = \hat U (\shortminus t)\hat\sigma _z\hat U (t)\Ket{\psi_{2}}
    =\sigma^{z}_{j}(t)\Ket{\psi_{2}}$
    
    $\hphantom{\Ket{\psi_{3}}}=\sigma^{z}_{j}(t)\frac{1 \pm \hat \sigma^{x}_{i}}{\sqrt{2}}\Ket{\psi}$
\\\\ 4.& Measure $\hat \sigma^{x}_{i}$.
    & $\comm_{\pm} = \Bra{\psi_{3}}\hat \sigma^{x}_{i}\Ket{\psi_{3}}$
    
    $\hphantom{\comm_{\pm}}= \Bra{\psi}\frac{1 \pm \hat \sigma^{x}_{i}}{\sqrt{2}} \hat \sigma^{z}_{j}(t) \hat \sigma^{x}_{i} \hat \sigma^{z}_{j}(t) \frac{1 \pm \hat \sigma^{x}_{i}}{\sqrt{2}}\Ket{\psi}$
\end{tabular}

\subsection{Realizing $1\pm \hat \sigma^{x}$ via the $\Ket{\rm f}$ state}

The OTOC defined in Eq.~\ref{eq:otocAppendix} can be extracted for an arbitrary initial state $\Ket{\psi}$ by expanding the non-unitary operation $(1\pm \hat \sigma^{x})/\sqrt{2}$ into a unitary operator in the three-level manifold of the transmon, including the second excited state $\Ket{\rm f}$. Here, we can realize $\comm_\pm$ with the sequence as follows:

\begin{tabular}{p{8pt}p{0.5\textwidth}@{\hspace{12pt}}p{0.4\textwidth}}
\\ 1. & Prepare the arbitrary initial state $\Ket{\psi}$ for all lattice sites in the qubit manifold.& 
\\\\ 2. & (a) Apply the rotation $R_i^y(\pm\frac{\pi}{2})$ to qubit $i$ (in its two-level manifold),
    & $\Ket{\psi_{2a}} =\left[\Ket{\rm g}\frac{\Bra{\rm g}\pm\Bra{\rm e}}{\sqrt{2}}+ \Ket{\rm e}\frac{\Bra{\rm e}\mp\Bra{\rm g}}{\sqrt{2}} + \Ket{\rm f}\Bra{\rm f}\right]_{i}\Ket{\psi}$
    
    $ \hphantom{\Ket{\psi_{2a}}} =\left[\Ket{\rm g}\frac{\Bra{\rm g}\pm\Bra{\rm e}}{\sqrt{2}}+ \Ket{\rm e}\frac{\Bra{\rm e}\mp\Bra{\rm g}}{\sqrt{2}}\right]_{i}\Ket{\psi}$
\\\\ & (b) Apply a $\pi$-pulse to qubit $i$ at the $\Ket{\mathrm{e}}\leftrightarrow\Ket{\mathrm{f}}$ transition frequency. Note that due to the large anharmonicity, all subsequent gates in the qubit manifold conserve the $\Ket{\rm f}$-level population.
    & $\Ket{\psi_{2b}} =
    \left[\Ket{\rm f}\Bra{\rm e} + \Ket{\rm e}\Bra{\rm f} + \Ket{\rm g}\Bra{\rm g}\right]_{i}\ket{\psi_{2a}} $
    
    $\hphantom{\Ket{\psi_{2b}}} = \left[\Ket{\rm g}\frac{\Bra{\rm g}\pm\Bra{\rm e}}{\sqrt{2}}+ \Ket{\rm f}\frac{\Bra{\rm e}\mp\Bra{\rm g}}{\sqrt{2}}\right]_{i}\Ket{\psi}$
\\\\ & (c) Apply the rotation  $R_i^y(\mp\frac{\pi}{2})$ to qubit $i$,
    & $\Ket{\psi_{2c}} =\left[\Ket{\rm g}\frac{\Bra{\rm g}\mp\Bra{\rm e}}{\sqrt{2}}+ \Ket{\rm e}\frac{\Bra{\rm e}\pm\Bra{\rm g}}{\sqrt{2}} + \Ket{\rm f}\Bra{\rm f}\right]_{i}\Ket{\psi_{2b}}$
    
    $\hphantom{\Ket{\psi_{2c}}} = \left[\frac{\Ket{\rm g}\pm\Bra{\rm e}}{\sqrt{2}}\frac{\Bra{\rm g}\pm\Bra{\rm e}}{\sqrt{2}}+ \Ket{\rm f}\frac{\Bra{\rm e}\mp\Bra{\rm g}}{\sqrt{2}}\right]_{i}\Ket{\psi}$
    
    $\hphantom{\Ket{\psi_{2c}}} = \frac{1 \pm \hat\sigma^{x}_{i}}{2}\Ket{\psi} + \left[\Ket{\rm f}\frac{\Bra{\rm e}\mp\Bra{\rm g}}{\sqrt{2}}\right]_{i}\Ket{\psi}$
\\\\ 3. & Apply $\sigma^{z}_{j}(t)$, as above.
    & $\Ket{\psi_{3}} =\sigma^{z}_{j}(t)\Ket{\psi_{2c}}$
\\\\ 4.& (a) Apply the rotation $R_i^y(-\frac{\pi}{2})$ to qubit $i$, & $\Ket{\psi_{4}} =R_i^y(\frac{-\pi}{2})\Ket{\psi_{3}}$
\\\\ & (b) Measure the occupation of \emph{all} qubits. If any are at $\Ket{\rm f}$, assign $\comm_{\pm} = 0$; otherwise assign $\comm_{\pm} = 2\sigma^{z}_{i}$.
    & $\comm_{\pm} = \Bra{\psi_{4}}\hat P_{\rm ge} 2\hat \sigma^z_{i}\hat P_{\rm ge}\Ket{\psi_{4}}$
\end{tabular}

\vspace{10pt}
The condition of finding the states of all lattice sites in the qubit manifold in the final step can be achieved with the projection operator into the qubit subspace,
\begin{equation}
\hat P_{\rm ge} = \prod_k\left[\Ket{\rm g}_k\Bra{\rm g}_k + \Ket{\rm e}_k\Bra{\rm e}_k\right].
\end{equation}
Due to the large anharmonicity, all gates applied to the qubit manifold do not mix the populations in the qubit manifold and in the $\Ket{\rm f}$ level, and therefore $\hat P_{\rm ge}$ commutes with all gates applied to the qubit manifold. Therefore, one finds
\begin{equation*}\begin{split}
    \hat P_{\rm ge}\ket{\psi_{4}} & = \hat P_{\rm ge}R_i^y(-\tfrac{\pi}{2})\sigma^{z}_{j}(t)\frac{1 \pm \hat \sigma^{x}_{i}}{2}\Ket{\psi} + \hat P_{\rm ge}R_i^y(-\tfrac{\pi}{2})\sigma^{z}_{j}(t)\left[\Ket{\rm f}\frac{\Bra{\rm e}\mp\Bra{\rm g}}{\sqrt{2}}\right]_{i}\Ket{\psi}
    \\ & = R_i^y(-\tfrac{\pi}{2})\sigma^{z}_{j}(t)\frac{1 \pm \hat \sigma^{x}_{i}}{2}\hat P_{\rm ge}\Ket{\psi} + R_i^y(-\tfrac{\pi}{2})\sigma^{z}_{j}(t)\hat P_{\rm ge}\left[\Ket{\rm f}\frac{\Bra{\rm e}\mp\Bra{\rm g}}{\sqrt{2}}\right]_{i}\Ket{\psi}
    = R_i^y(-\tfrac{\pi}{2})\sigma^{z}_{j}(t)\frac{1 \pm \hat \sigma^{x}_{i}}{2}\Ket{\psi},
\end{split}\end{equation*}
and so we have
\begin{equation*}\begin{split}
    \comm_{\pm} & = 2\Bra{\psi_{4}}\hat P_{\rm ge} \hat \sigma^{z}_{i}\hat P_{\rm ge}\Ket{\psi_{4}} 
    \\ &= 2\Bra{\psi}\frac{1 \pm \hat \sigma^{x}_{i}}{2}\hat \sigma^{z}_{j}(t) R_i^y(\tfrac{\pi}{2}) \hat \sigma^z_{i}R_i^y(-\tfrac{\pi}{2})\sigma^{z}_{j}(t)\frac{1 \pm \hat \sigma^{x}_{i}}{2}\Ket{\psi} 
    \\ & = 2\Bra{\psi}\frac{1 \pm \hat \sigma^{x}_{i}}{2}\hat \sigma^{z}_{j}(t)  \hat \sigma^x_{i}\sigma^{z}_{j}(t)\frac{1 \pm \hat \sigma^{x}_{i}}{2}\Ket{\psi} 
    \\ & =\Bra{\psi}\frac{1 \pm \hat \sigma^{x}_{i}}{\sqrt{2}} \hat \sigma^{z}_{j}(t) \hat \sigma^{x}_{i} \hat \sigma^{z}_{j}(t) \frac{1 \pm \hat \sigma^{x}_{i}}{\sqrt{2}}\Ket{\psi},
\end{split}\end{equation*}
as intended.

\section{Pulse sequences}

\subsection{Realization and calibration of the time-reversed unitary evolution}

\begin{figure*}
\includegraphics{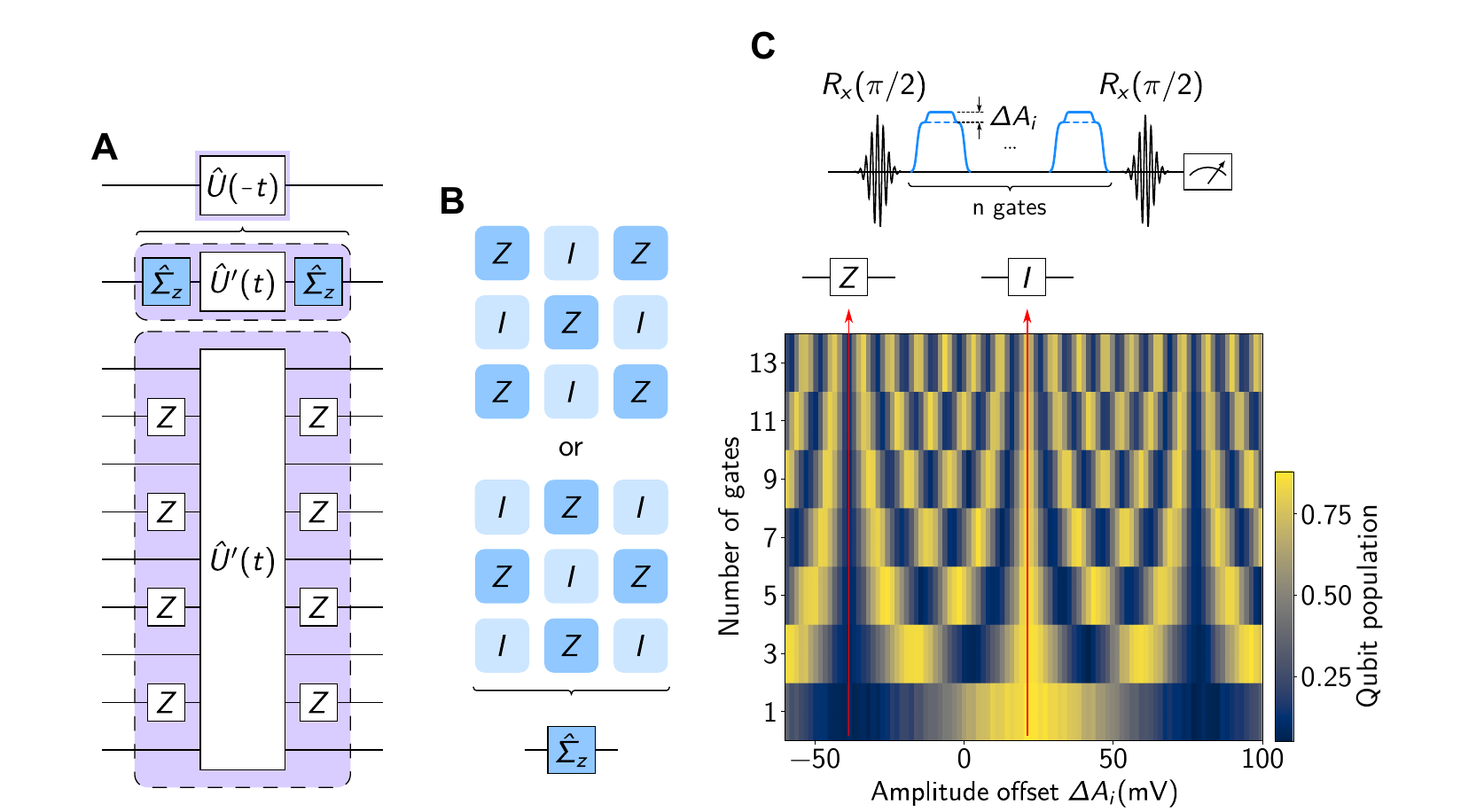}
\caption{
\textbf{Realization and calibration of $\SZ$}
(\textbf{A}) Circuit diagram for the realization of $\hat U(\shortminus t)$.
(\textbf{B}) The two possibilities for realizing $\SZ$, by applying Pauli-$Z$ gates to every other qubit in the lattice in both dimensions, while the remaining qubits receive an identity gate $I$.
(\textbf{C}) Calibration of the Pauli-$Z$ gates and identity gates $I$ for the $\SZ$ operation. The pulse form is schematically shown at the top: we enclose an odd number $n$ of detuning pulses with amplitudes $A_i+\Delta A_i$ within a Ramsey sequence. Based on the final qubit population we can infer the amplitude offset conditions for the $Z$ and $I$ gate, respectively.
}
\label{fig:zgates}
\end{figure*}

One of the key elements in our pulse sequences is the time-reversed unitary evolution $\hat U (\shortminus t)$. Its circuit diagram is given in Fig.~\ref{fig:zgates}A. We sandwich a forward time evolution step $\hat U \pr(t)$, under the Hamiltonian $\hat H \pr$, between two sets of single qubit gates $\SZ=\Pi_{i\in\mathrm{black}} \hat\sigma _i ^z$, which are Pauli-$Z$ gates applied to every other qubit in the lattice in both dimensions. The two possible experimental realizations of $\SZ$ are shown in Fig.~\ref{fig:zgates}B, where $Z$ denotes the Pauli $Z$-gate (which is identical to a $R^z(\pi)$ rotation up to a global phase), and $I$ denotes the identity gate. The Hamiltonian $\hat H \pr$ is identical to $\hat H$, but with flipped disorder frequencies, $\Delta\omega_i\rightarrow\shortminus\Delta\omega_i$. In the hard-core Bose-Hubbard model (Eq.~3), we find $\SZ\hat \sigma^+_i\hat\sigma^-_j\SZ=-\sigma^{+}_i\hat\sigma^{-}_j$ for any pair of adjacent qubits $\langle i,j \rangle$ and therefore
\begin{equation}
\SZ\hat H \pr\SZ/\hbar=\sum_{\langle i,j\rangle}J_{ij}\hat\sigma _i^+ \hat\sigma _j +\sum_i\frac{\shortminus\Delta\omega_i}2\hat\sigma _i^z=-\hat H/\hbar.
\end{equation}
The natural time evolution now yields
\begin{equation}
\SZ\hat U \pr(t)\SZ=\SZ e^{-i\hat H \pr t}\SZ=e^{\shortminus i(\shortminus\hat H)t}=\hat U (\shortminus t),
\end{equation}
realizing the required time-reversed evolution under $\hat H$. As we show in Fig.~\ref{fig:otocseq}, we apply the $\SZ$ operation as virtual $Z$ gates~\cite{McKay2017}, by redefining the coordinate system of the successive tomography pulses. The initial $\SZ$ operation at the beginning of the reverse time evolution step $\hat U(\shortminus t)$ must be executed as physical gates, since the relative single-qubit phases are relevant for the subsequent unitary time evolution step.

Figure~\ref{fig:zgates}C show an example of a calibration measurement that we have performed for calibrating both the $Z$ and $I$ gates. The fast flux $Z$ control pulses in between forward and backward time evolution steps serve a double purpose. On the one hand, they are used to switch unitary evolution blocks on and off: if the qubits are tuned on or close to resonance, then unitary time evolution leads to the analog block $U(\pm t)$. In order to freeze the lattice dynamics, we mutually detune the qubits, such that neighboring qubits are detuned by $\gtrsim\SI{300}{MHz}$ in order to efficiently suppress time evolution. The effective interaction strength for neighboring qubits detuned by $\Delta/2\pi=\SI{300}{MHz}$ is
\begin{equation}
J_{\mathrm{freeze}}=\frac{-|\Delta|+\sqrt{\Delta^2 + 4 J^2}}{2}\approx 2\pi\times\SI{0.2}{MHz}.
\end{equation}

On the other hand, the fast flux $Z$ control pulses are used to achieve the correct $Z$ rotations required for realizing the first $\SZ$ prior to the unitary time evolution step. In Fig.~\ref{fig:zgates}C, we demonstrate the calibration procedure used for each qubit separately. On top of the main detuning pulse (blue in  the calibration sequence of Fig.~\ref{fig:zgates}C), we add a small additional detuning pulse of amplitude $\Delta A _i$. By sweeping $\Delta A _i$ and enclosing a varying number of $Z$ gates in a Ramsey sequence, we find the conditions for realizing a $Z$ gate and an $I$ gate, respectively, see the the red lines in Fig.~\ref{fig:zgates}C. In order to minimize any distorting effects due to the ac-Stark push of the other qubits, we bias the other qubits to the frequencies where they are parked in experiment while the unitary evolution is frozen (except for qubit $i$, which we have to tune away from the common resonance frequency in order to not interfere with the calibration).

We further optimize the calibrated $Z$ gates with a Loschmidt echo sequence, minimizing the deviation from the initially prepared state.

\subsection{Detailed pulse sequence for OTOC experiment}

\begin{figure*}
\includegraphics{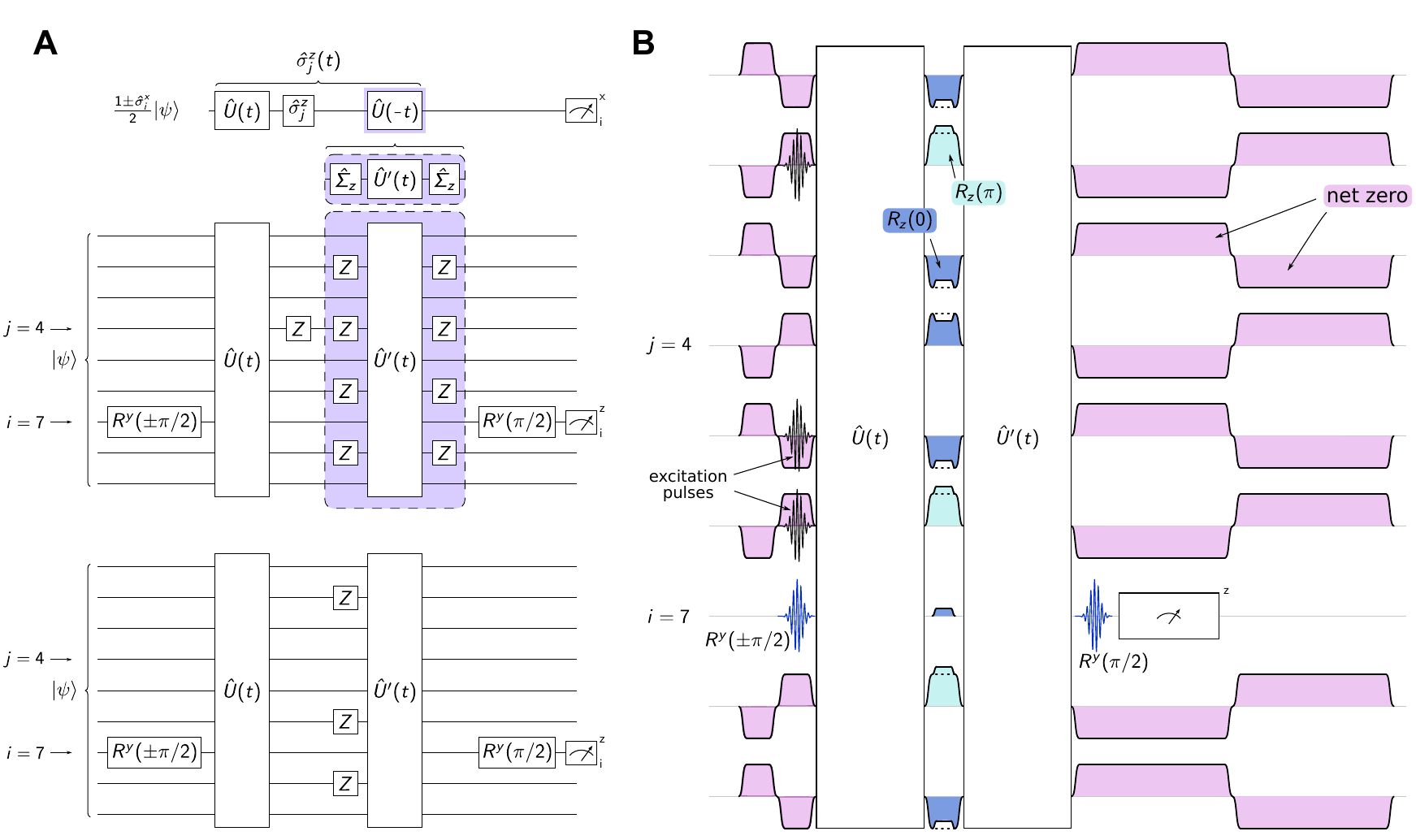}
\caption{
\textbf{Pulse sequence used in OTOC experiments}
(\textbf{A}) Schematic pulse sequence and circuit diagram for the OTOC experiments. The circuit diagram at the bottom shows the gates and unitary time evolution blocks that are physically applied in experiment.
(\textbf{B}) Detailed pulse sequence used in experiments. Qubit $i$, the qubit that receives the initial perturbation and is read out at the end of the OTOC pulse sequence remains at the common qubit frequency $\wq$ throughout the pulse sequence. The remaining qubits are tuned out of resonance outside of the time evolution blocks.
}
\label{fig:otocseq}
\end{figure*}

Figure~\ref{fig:otocseq}A shows the schematic pulse sequence used in the OTOC experiments, as well as corresponding circuit diagrams. In this example, qubit $i$ receives the initial preparation pulse. As detailed above, we realize the operation $(1\pm\hat\sigma _i^x)/\sqrt{2}$ with the rotation gate $R^y_i(\pm\pi/2)$ after preparing the initial state $\Ket\psi=\Ket{\mathrm{g}}_i\otimes\prod_{k\neq i}\Ket\psi _k$, where qubit $i$ is initially restricted to be in its groundstate $\Ket{\mathrm{g}}_i$. Depending on the lattice site $j$, where we apply the butterfly operator $\hat\sigma _j^z$, we can merge the butterfly $Z$ gate and the $Z$ gate as part of $\SZ$ according to $ZZ=I$. The final $\SZ$ operation is either applied virtually, by absorbing the $Z$ rotation into the final $R^y(\pm\pi/2)$ rotation, or omitted since we only require information on the state of qubit $i$.

A more detailed pulse sequence for the OTOC experiments is given in Fig.~\ref{fig:otocseq}B. In order to freeze the system dynamics most efficiently, we detune the qubits from each other according to a checkerboard pattern, such than adjacent sites are tuned in different directions. The exception is qubit $i$, which remains at the common rotating frame frequency $\wq$ throughout the entire pulse sequence in order to lock its reference phase. For initial state preparation and qubit readout, we use net-zero $Z$ control pulses~\cite{Rol2019} in order to avoid any long-timescale transient distortions. Pulse amplitude corrections $\Delta A _i$ are visible in the central freeze-out and $\SZ$ operation, resulting in alternating phase rotation gates $R^z(\pi)$ and $R^z(0)\equiv I$.

In the presence of frequency disorder, we invert the rotating frame disorder frequencies $\Delta\omega_i$ during $\hat U (t\pr)$, as compared to the configuration during $\hat U (t)$. By observing a Loschmidt echo for a disordered lattice we have confirmed that this results in an efficient time reversal.

\subsection{Minimization of conditional phase errors}

\begin{figure}
    \includegraphics[width=0.5\textwidth]{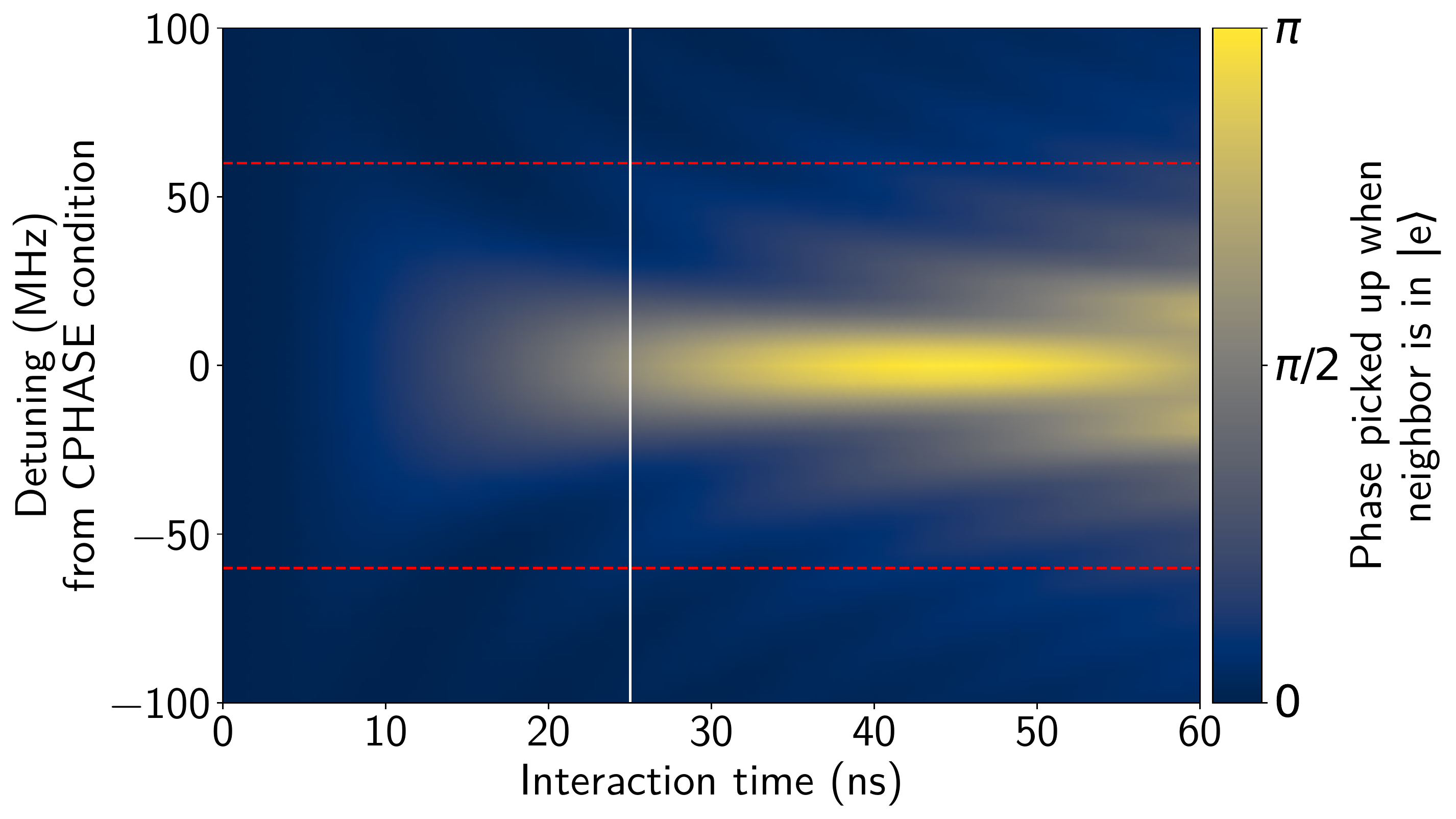}
    \caption{\textbf{Simulation of the conditional phase accumulation} We show a simulation of the conditional phase (CPHASE) picked up when a neighboring qubit is in $\Ket{\mathrm{e}}$ instead of in $\Ket{\mathrm{g}}$ as a function of the detuning form the CPHASE condition (where the two-qubit states $\Ket{\mathrm{gf}}$ and $\Ket{\mathrm{ee}}$ are degenerate) and the interaction time. The white line show the duration of the interaction time in our experiment, which corresponds to the time between the two unitary time evolution blocks in Fig.~\ref{fig:otocseq}B.}
    \label{fig:cphase_accumulation}
\end{figure}

The fast flux pulses applied between the two unitary evolution steps of the OTOC pulse sequence (Fig.~\ref{fig:otocseq}) freeze the system dynamics and apply the Pauli-$Z$ gates and identity gates $I$ required for realizing $\hat U (\shortminus t)$ and the butterfly operation. In this scheme, one possible error mode is the conditional phase (CPHASE) accumulated on the $\ket{\mathrm{ee}}$ state of two neighboring qubits as a result of the interaction with the $\ket{\mathrm{gf}}$ states. This is the interaction used to realize a CPHASE gate~\cite{Krantz2019}, and is maximized when the states $\ket{\mathrm{ee}}$ and $\ket{\mathrm{gf}}$ are degenerate, i.e. when
\begin{equation}
\omega_{01}^{\mathrm{q1}}+\omega_{01}^{\mathrm{q2}}=\omega_{01}^{\mathrm{q1}}+\omega_{12}^{\mathrm{q1}}
\end{equation}
and therefore
\begin{equation}
\omega_{12}^{\mathrm{q1}}=\omega_{01}^{\mathrm{q1}}+U=\omega_{01}^{\mathrm{q2}},
\end{equation}
where $\omega_{ij}^{\mathrm{q}}$ is the transition frequency of qubit $\mathrm{q}$ between levels $i$ and $j$, and $U$ is the qubit anharmonicity.

A simulation of this conditional phase accrual is shown in Fig.~\ref{fig:cphase_accumulation}. We suppress this error mode by detuning neighboring qubits by at least $\SI{300}{MHz}$, such that we maintain a $\SI{60}{MHz}$ clearance from the CPHASE condition (red dashed lines).

\section{Notes on data evaluation}

\subsection{Loschmidt echo}

The extracted fidelities and concurrence data at the end of the Loschmidt echo sequence (Fig.~2B) have a small dependence on the phase $\phi$ of the prepared entangled state $\Ket{\Psi^{74}}$. The reason is that the Bell states ($\phi=0,\pi$) are eigenstates of $\mathrm{iSWAP}$ and are therefore robust against bringing qubits on resonance (or sweeping qubits through their common resonance). In contrast, the two-qubit states with $\phi=\pm\pi/2$ are not eigenstates of $\mathrm{iSWAP}$ and therefore udnergo free evolution when qubits are tuned on resonance. This effect leads to the slight oscillations visible in fidelity and concurrence data in Fig.~2B. We have chosen the preparation phase for the prepared state (Fig.~2C) to yield maximum fidelity and are showing data for the Loschmidt echo with echo time $t>0$ and in Fig.~2C for a constant preparation phase $\phi$ where fluctuations are least dominant and overall fidelities are best.

We extract the dephasing limit (dashed line in Fig.~2B) via first calculating the density matrix of the prepared state $\hat\rho _0$ subject to a net dephasing of the entire lattice with $T_{2,\mathrm{eff}}=\SI{0.884}{\micro s}$ (see Sec.~\ref{sec:dephasing}) according to
\begin{equation}
\hat\rho _{0,\mathrm{deph}} (t)=e^{-2t/T_{2,\mathrm{eff}}}\hat\rho _0+\left(1-e^{-2t/T_{2,\mathrm{eff}}}\right)\hat\rho _\infty^{(2\mathrm{q})},
\end{equation}
where $t$ is the Loschmidt echo time (therefore the factor of $2$ in the exponent), $\hat\rho _0$ is the density matrix of the prepared state $\Ket{\Psi^{74}}$ in the two-qubit subsystem of qubits $7$ and $4$, and $\hat\rho_\infty^{(2\mathrm{q})}=\tr_{\{i\notin 4,7\}} \hat\rho _\infty$, where $\hat\rho _\infty$ is the completely random classical state of the nine-qubit system containing one excitation.

We have verified the validity of our simple model for dephasing by comparing to numerical simulations of the Lindblad master equation, assuming individual dephasing rates for all qubits.

\subsubsection{Single-qubit phase accrual in the two-qubit density matrix}

Tuning qubit frequencies away from the common reference frequency $\wq$ causes a local phase accrual in each qubit's subspace. The effect of this phase is minimized by fixing the qubit that receives the initial perturbation (qubit seven) at the common reference frequency $\wq$ throughout the pulse sequence and detuning the other qubits to freeze out the lattice dynamics and realize backward time evolution. However, the accumulated phase needs to be taken into account when reconstructing the density matrix for the  two-qubit subsystem consisting of qubits seven and four, since qubit four leaves the reference frequency during the sequence. The accumulated phase can by compensated in two ways: the first method involves applying a virtual $R^z(\varphi)$ gate prior to applying the single-qubit rotations used for state tomography. Alternatively, we can apply the $R^z(\varphi)$ rotation to qubit four in post-processing after reconstructing the two-qubit density matrix. In this second scheme, the spin of qubit four is measured along the rotated $X$ and $Y$ axes, $X^\prime$ and $Y^\prime$. Using these measurements, the reconstructed density matrix is
\begin{equation}
    \hat{\rho}_{4,7}^\prime = \sum_i p_i \ket{\psi_i^\prime} \bra{\psi_i^\prime},
\end{equation}
where $\ket{\psi_i^\prime}=R^z_{4}(\varphi)\ket{\psi_i}$ and $\ket{\psi_i}$ is the quantum state in the logical frame of both qubits. $\hat{\rho}_{4,7}^\prime$ is then transformed to the density matrix in the logical frame by applying a unitary transformation in post-processing:
\begin{equation*}
    \hat{\rho}_{4,7} = R^z_{4}(-\varphi)  \: \hat{\rho}_{4,7}^\prime \: R^z_{4}(\varphi).
\end{equation*}
We find the correct phase $\varphi$ for the results in Fig.~2 via the second described method in post-processing.

\subsubsection{Concurrence}

Concurrence is an entanglement monotone used to measure the entanglement between two qubits, ranging from $0$ to $1$. For a given two-qubit density matrix $\hat\rho$, the concurrence $\mathcal{C}(\hat\rho)$ is defined as
\begin{equation}
    \mathcal{C}(\hat\rho)=\text{max}(0,\,\lambda_1-\lambda_2-\lambda_3-\lambda_4)
\end{equation}
where $\lambda_1 \geq \lambda_{2}\geq \lambda_{3}\geq \lambda_{4}$ are the eigenvalues of the Hermitian matrix $\hat R=\sqrt{\sqrt{\hat\rho}\tilde\rho\sqrt{\hat\rho}}$, measuring the overlap between $\hat\rho$ and $\tilde{\rho}=(\hat\sigma ^y \otimes \hat\sigma ^y)\hat\rho ^*(\hat\sigma ^y \otimes \hat\sigma ^y)$, the result of a spin flip transformation on the original density-matrix $\hat\rho$. Using this framework, the entanglement of formation $E(\hat\rho)$ of a bipartite mixed state described by $\hat\rho$ can be related to concurrence~\cite{Hill1997}:
\begin{equation}
    E(\hat\rho)=h\left(\frac{1+\sqrt{1-C(\hat\rho)^2}}{2}\right)
\end{equation}
where $h(x)=-x\:\rm{log}(x)-(1-x)\:\rm{log}(1-x)$ is the binary entropy function.

\subsection{OTOC data}

To obtain the OTOC data that we show in Fig.~3,~4, we sum the measured quantities $\comm_\pm$ from all measurements where the butterfly operator $\hat W _j(t)$ is applied at the same Manhattan distance $j$. We account for small baseline shifts -- which accumulate in this sum -- by subtracting the baseline offset in $\comm_\pm$ for each Manhattan distance individually.

We perform numerical simulations using QuTiP~\cite{Johansson2013}. To account for the qubit frequency dependence of the coupling strengths between qubits $i$ and $j$, we scale the measured coupling strengths at qubit frequencies of $\SI{5.5}{GHz}$ according to $J_{ij}(\omega_i,\omega_j)\propto\sqrt{\omega_1\omega_2}$. The reference operation frequency $\wq$ in Fig.~3,~4 is $\SI{5.3}{GHz}$.

\subsection{Compensating for dephasing in OTOC measurements}\label{sec:dephasing}

\begin{figure}
\includegraphics{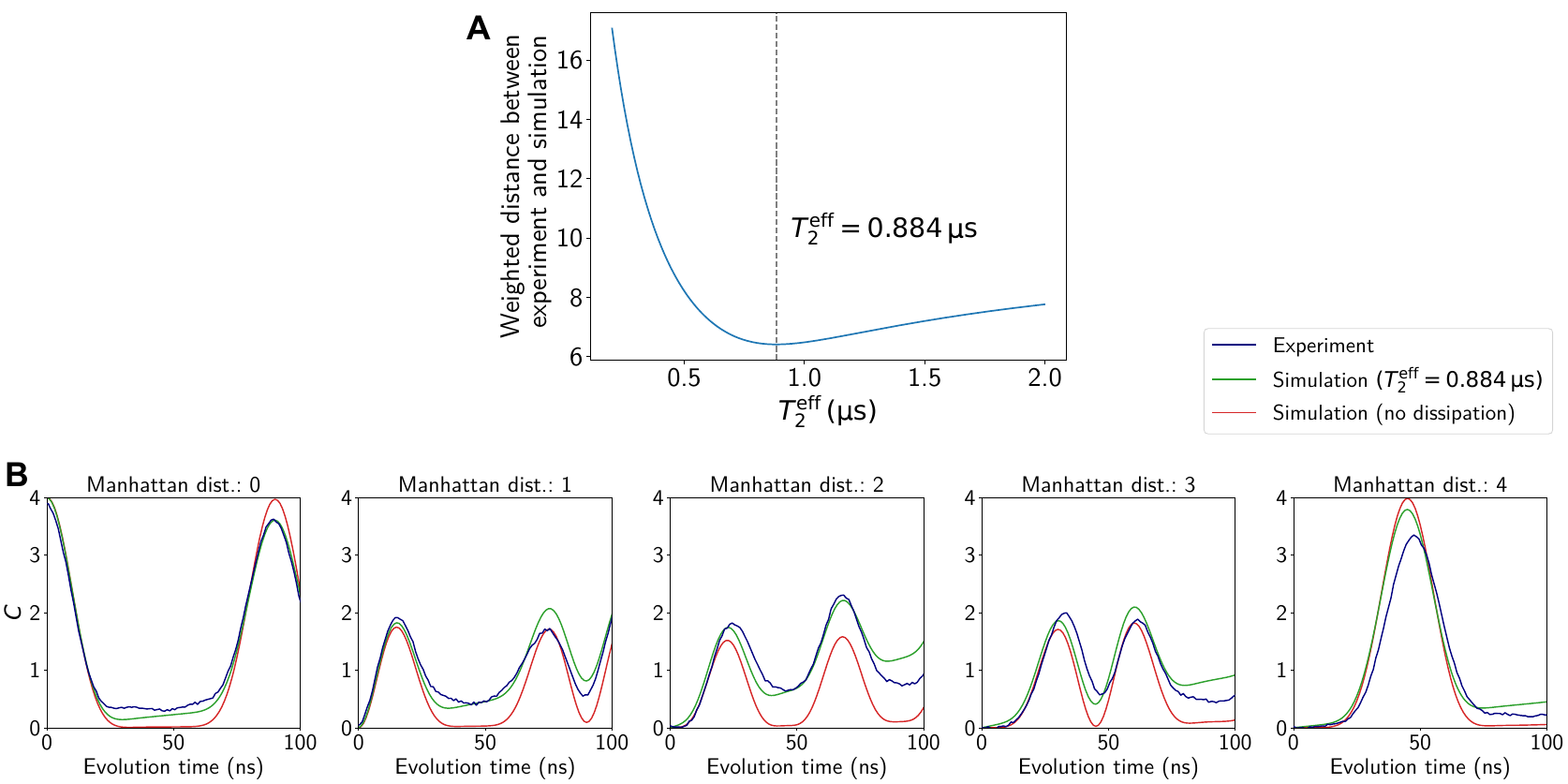}
\caption{\textbf{Extracting the effective dephasing time from measured OTOC data for the no-disorder scenario}
(\textbf{A}) We plot the weighted and interpolated distance between the experimental OTOC data (without disorder) and the respective numerical simulation of the experimental Hamiltonian with varying strengths of qubit dephasing. We plot the effective net dephaing time $T_2^{\rm eff}$.
(\textbf{B}) Experimental data (blue), numerical simulations based on the experimental Hamiltonian including the third transmon levels and the calculated non-nearest neighbor couplings as well as the extracted qubit dephasing (green), and the numerical simulation based on the ideal hard-core Bose-Hubbard model (Eq.~3, red).
}
\label{fig:findT2}
\end{figure}

As described above, our experimental system is subject to decoherence. While the measured $T_1$ times of the qubits are much larger than our time evolution such that energy relaxation does not significantly impact our results, the shorter dephasing ($T_2$) times are noticeable. It has been noted in Ref.~\cite{Swingle2018a} that measured OTOC data can be corrected for dephasing errors.

We calculate the effects of dephasing on our system using a simple model. We assume that the number of excitations in the system, $\hat n_{\rm ex} = \sum_{i}\frac{\hat \sigma^z_i+1}{2}$, is conserved, but within each $\hat n_{\rm ex} = n_{\rm ex}$ sector, dephasing drives the system towards an infinite temperature state where all configurations are equally likely. According to the model, the density matrix of the system subject to dephasing is
\begin{equation}
    \rho(t) = e^{-t/T_2^{\rm eff}}\rho^{\rm ideal}(t) + \left(1-e^{-t/T_2^{\rm eff}}\right)\sum_{n_{\rm ex}}P_{n_{\rm ex}}(t)\rho_{\infty,n_{\rm ex}}
    \label{eq:dephasing}
\end{equation}
where $\rho^{\rm ideal}(t)$ is the dephasing-free density matrix, $T_2^{\rm eff}$ is the effective net dephasing time, $P_{n_{\rm ex}}(t)$ is the probability of finding $n_{\rm ex}$ excitations in the system, and $\rho_{\infty,n_{\rm ex}}$ is the corresponding infinite temperature state. In Eq.~\ref{eq:dephasing}, $t$ denotes the total evolution time during an experiment.

As described in Sec.~\ref{sec:OTOCconstruction}, our procedure for obtaining the OTOCs uses a final measurement of $\hat \sigma^x_i$. In the infinite temperature state, we find
\begin{equation}
    \langle\hat \sigma^x_i\rangle_{\infty} = \tr[\hat\rho_{\infty}\hat \sigma^x_i] = 0
\end{equation}
for any $n_{\rm ex}$ and the measured quantities $\comm_{\pm}^{\rm meas}$ of Eq.~\ref{eq:Cpmdef} behave simply as
\begin{equation}
    \comm_{\pm}^{\rm meas}(t) = e^{-2t/T_2^{\rm eff}}\comm_{\pm}^{\rm ideal}(t).
\end{equation}
The factor of two in the exponent comes from the fact that the OTOC measurement process involves two time evolution steps of length $t$. We can therefore account for the effects of dephasing by assigning
\begin{equation}
    \comm_{\pm}(t) = e^{2t/T_2^{\rm eff}}\comm_{\pm}^{\rm meas}(t).
\end{equation}

We evaluate $T_2^{\rm eff}$ by comparing the numerical simulation of the no-disorder, zero-excitation OTOC to the respective experimental results and by minimizing their difference, see Fig.~\ref{fig:findT2}. We find an effective $T_2^{\rm eff} = \SI{0.884}{\micro s}$, which is a net (aggregate) value from all nine qubits.

\subsection{Extracting the light cone from OTOC data of disordered lattices}

\begin{figure}
\includegraphics{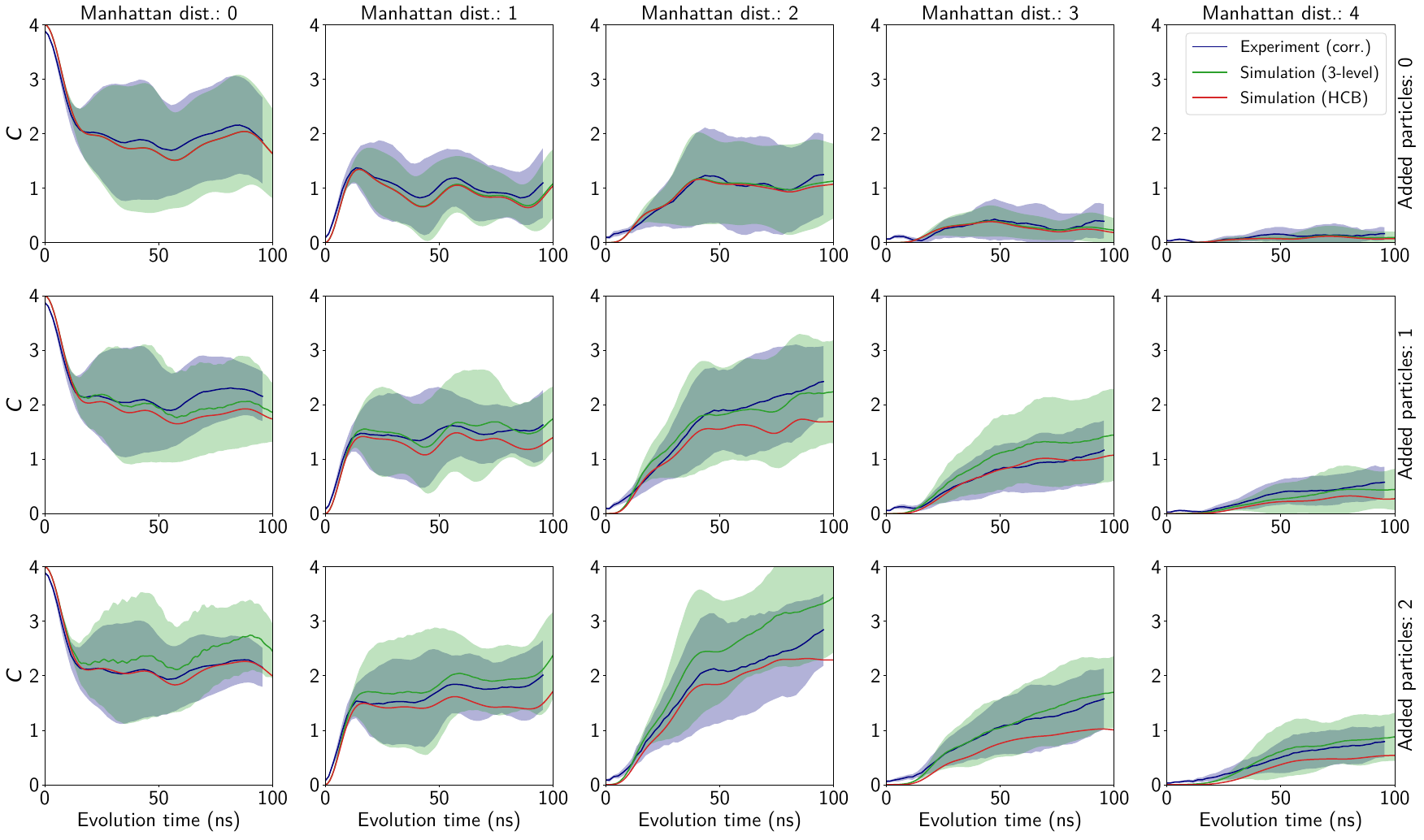}
\caption{\textbf{Overview of OTOC data from disordered lattices}
Experimental data (blue) and numerical simulations for OTOC data based on our experimental Hamiltonian including the third transmon levels and the calculated non-nearest neighbor couplings (green).  Experimental data are modified to account for the effects of decoherence with $T_2^{\rm eff} = \SI{0.884}{\micro s}$. Shaded regions show the standard deviation of the data for the twelve random disorder realizations, while the solid lines represent the average. The red line corresponds to the numerical simulation based on the ideal hard-core Bose-Hubbard model (Eq.~3). While corrected experimental data and the realistic numerical simulation show good agreement, we observe more pronounced deviations between experiment and the ideal simulation, mainly caused by the always-on $ZZ$ interaction, see Sec.~\ref{sec:ZZ}.
}
\label{fig:overview}
\end{figure}

\begin{figure}
\includegraphics{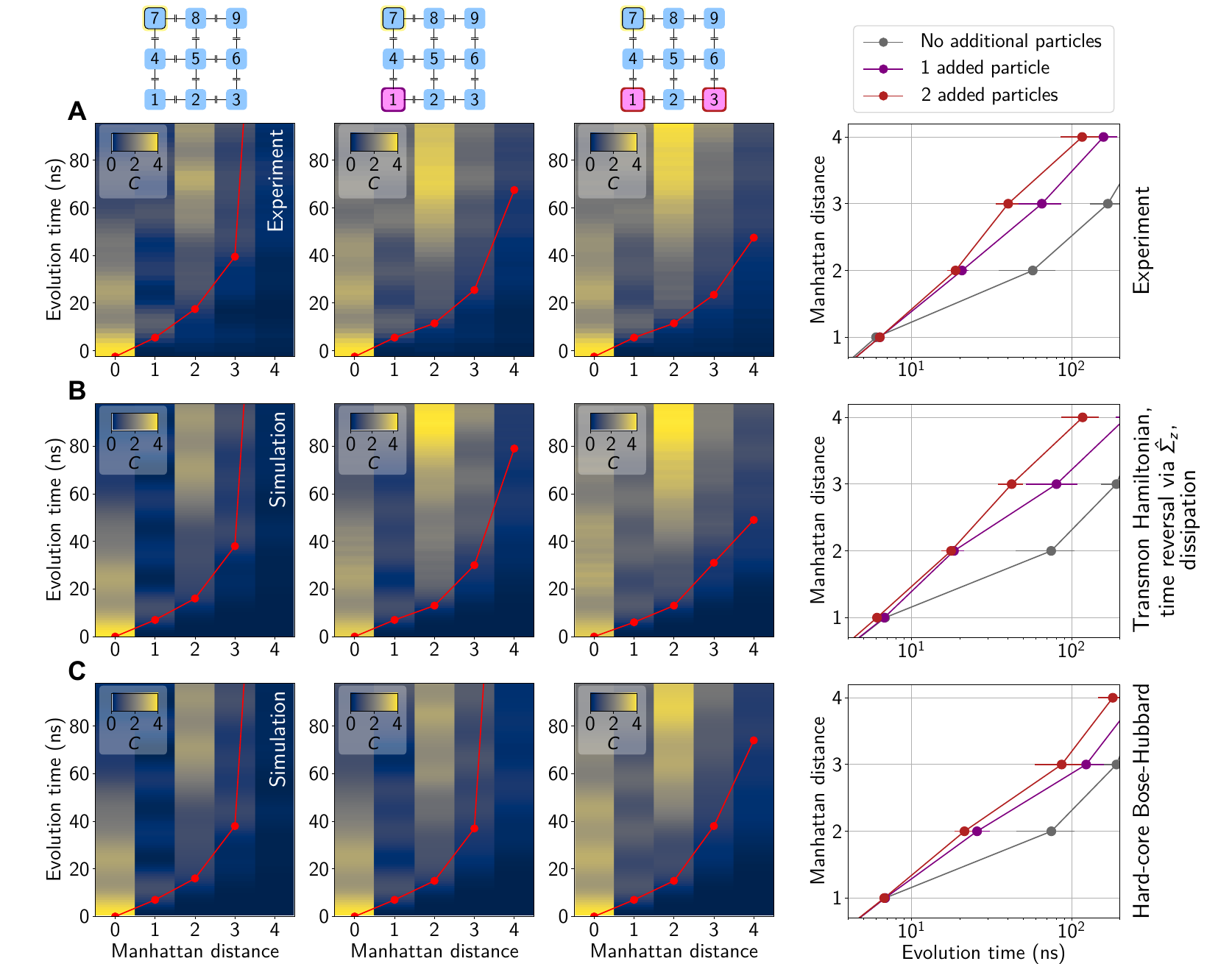}
\caption{\textbf{OTOC measurements in disordered lattice in comparison with different numerical simulations}
(\textbf{A}) Representative OTOC data in a disordered lattice with $\langle\langle(\Delta\omega _i)^2\rangle\rangle^{1/2}=2.7J$ and a varying number of particles in the lattice, according to Fig.~4A (compensated for dephasing).
(\textbf{B}) Numerical simulation for the same disorder realization and based on our experimental Hamiltonian, including the third transmon levels and the calculated non-nearest neighbor couplings.
(\textbf{C}) Numerical simulation for the same disorder realization based on the hard-core Bose-Hubbard model (Eq.~3).
}
\label{fig:otocdissimvar}
\end{figure}

Experimental data, with compensated effects of dephasing as described above, and numerical simulations of the OTOC data in the presence of disorder ($\langle\langle(\Delta\omega _i)^2\rangle\rangle^{1/2}=2.7J$) are given in Fig.~\ref{fig:overview}. We observe good agreement between experimental data (blue) and numerical simulations of the experimental Hamiltonian including the third transmon levels and the calculated non-nearest neighbor couplings.

In order to quantify the speed of interaction propagation in the disordered lattice, we extract a `light cone' from measured OTOC data, defined by the time when a threshold value of $0.6$ is first reached at a given Manhattan distance (red dots and lines in Fig.~4A). If the threshold is not reached before the maximum evolution time in our data set ($\SI{100}{ns}$), we assign a time of $\SI{300}{ns}$. We note that modifying the post-processing parameters does not alter our qualitative results.

In Fig.~\ref{fig:otocdissimvar}, we show OTOC data for the disordered lattice according to Fig.~4 in the main text. We compare experimental data (Fig.~\ref{fig:otocdissimvar}A) with a numerical simulation of our experimental Hamiltonian including the third transmon levels, a next-nearest neighbor coupling of $J/30$, and decoherence (Fig.~\ref{fig:otocdissimvar}B), and with a numerical simulation for the ideal hard-core Bose-Hubbard Hamiltonian (Fig.~\ref{fig:otocdissimvar}C). We observe good agreement between experimental data and the realistic simulation, with small deviations with respect to the ideal simulation. Experimental data and both versions of numerical simulations show consistent qualitative results.

\begin{figure}
\includegraphics{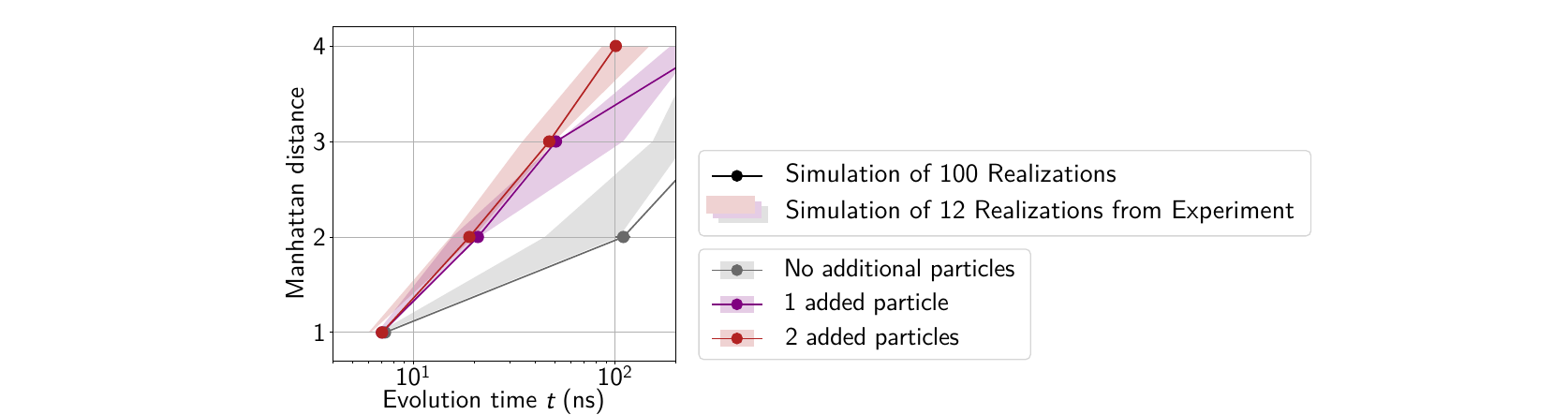}
\caption{\textbf{Numerical simulations of OTOC in disordered lattice}
We compare numerical simulations of the twelve random disorder realizations used in experiment (shaded areas) with $100$ additional random realizations of the same disorder strength $\langle\langle(\Delta\omega _i)^2\rangle\rangle^{1/2}=2.7J$. Their qualitative agreement confirms that the twelve realizations used in experiment are a representative set that show how localization is overcome in the presence of more particles in the lattice. The shaded regions show standard deviations of the means for the twelve realizations from experiment, while the data points show the average results of the median $20\%$ of simulation results.
}
\label{fig:otoc100}
\end{figure}

In addition, we demonstrate that the twelve random disorder realizations used in our experiment (Fig.~4) are a representative set by comparing numerical simulations of those twelve realizations with $100$ additional random realizations of the same disorder strength $\langle\langle(\Delta\omega _i)^2\rangle\rangle^{1/2}=2.7J$. The results in Fig.~\ref{fig:otoc100} are in qualitative agreement, showing that localization in the 2d hard-core Bose-Hubbard model is partially overcome with more particles present in the lattice. The small deviations between the two random sets is due to the unstable average of the extracted light cones, a consequence of the fact that the threshold value for some disorder realizations is never reached (in which case we assign an evolution time of $\SI{300}{ns}$). In order to suppress this effect, we compare the average of the twelve disorder realizations used in experiment with the average of the median $20\%$ of the additional $100$ disorder realizations in Fig.~\ref{fig:otoc100}.

\section{Limitations due to the transmon level structure\label{sec:ZZ}}

The most significant limitation in our experiment stems from the residual couplings between the qubits that are not captured by the two level Hamiltonian of Eq.~3. Even though the ratio $J/U\ll 1$ is sufficiently small such that the system can be described by a two-level Hamiltonian, the existence of the neglected $\Ket{\mathrm{f}}$ levels of the transmon qubits leads to a correction, known as the `always-on $ZZ$ interaction'~\cite{OMalley2015}. The Hamiltonian for the transmon system is
\begin{equation}
    \hat H_{\rm exp} = \hat H + \varepsilon \hat H_{\rm ZZ},
\end{equation}
where $\hat H$ is given in Eq.~3 and the correction is
\begin{equation}
\hat H_{\rm ZZ}/\hbar=-\sum_{\langle i,j\rangle}\frac{1+\hat\sigma _i^z\hat\sigma _j^z}{2}
\label{eq:HZZ}
\end{equation}
with its strength governed by
\begin{equation}
    \varepsilon = \frac{J^2}{|U|} \approx 2\pi\times\SI{0.54}{MHz}.
\end{equation}

The $ZZ$ terms are invariant under our time reversal procedure,
\begin{equation}
\SZ\hat H_{\rm exp}^{\prime}\SZ= \SZ(\hat H^{\prime}+\varepsilon \hat H_{\rm ZZ})\SZ = -\hat H + \varepsilon \hat H_{\rm ZZ}
\end{equation}
and so our time evolution operators are not exactly conjugate, but instead have
\begin{equation}
\hat U(t) = e^{-i(\hat H + \varepsilon\hat H_{\rm ZZ})t}, \qquad
\SZ\hat U \pr(t)\SZ=e^{i(\hat H - \varepsilon\hat H_{\rm ZZ})t}.
\end{equation}

We estimate the effect of these terms on our experimental results. Using a Trotter expansion~\cite{Suzuki1976}, we can rewrite
\begin{equation*}\begin{gathered}
\hat U(t) \approx e^{-i\hat Ht}\left(1 - i \varepsilon\int_{0}^{t}{\rm d}\tau \hat H_{\rm ZZ}(\tau)\right),
\\ \SZ\hat U \pr(t)\SZ \approx \left(1 - i \varepsilon\int_{0}^{t}{\rm d}\tau \hat H_{\rm ZZ}(\tau)\right)e^{i\hat Ht},
\\ \overline{\hat \sigma^z_j}(t) = \SZ\hat U \pr(t)\SZ\sigma^{z}_{j} \hat U(t) \approx
\hat \sigma^{z}_{j}(t) - i\varepsilon \int_{0}^{t}{\rm d}\tau \left\{\hat \sigma^{z}_{j}(t),\hat H_{\rm ZZ}(\tau)\right\}.
\end{gathered}\end{equation*}

\begin{table}
\centering
\caption{\textbf{Always-on $ZZ$ error estimates}. }
\vspace{8pt}
\label{tab:errorZZ}
{\renewcommand{\arraystretch}{1.7}   
\setlength{\tabcolsep}{10pt}
\begin{tabular}{c|cccccc|l}
$n_{\rm ex}$ & 0 & 1 & 2 & 3 & 4 & 5 \\ \hline 
\multirow{2}{*}{$T\left[\abs{\Delta\comm\approx\sqrt{\comm}}\right]$ (ns)}
    & $\infty$ & $\infty$ & 313 & 104 & 52 & 31 & $3\times3$ lattice\\ 
    & $\infty$ & $\infty$ & 467 & 156 & 78 & 47 & 7-qubit chain
\end{tabular}
}
\end{table}

We can then insert this result into the experimental formulation of Eq.~\ref{eq:CeqCmCp},~\ref{eq:Cpmdef}. To first order in $\varepsilon$, the measured OTOC is then given by
\begin{equation}\begin{split}
\comm^{\rm meas} & = 2+\comm_{-} - \comm_{+} \approx 
    2 - \Bra{\psi} \hat \sigma^{x}_{i}\overline{\hat \sigma^{z}_{j}}(t)^{\dagger} \hat \sigma^{x}_{i} \overline{\hat \sigma^{z}_{j}}(t) \Ket{\psi} - \Bra{\psi} \overline{\hat \sigma^{z}_{j}}(t)^{\dagger} \hat \sigma^{x}_{i} \overline{\hat \sigma^{z}_{j}}(t)\hat \sigma^{x}_{i} \Ket{\psi}
    \\ & = \Bra{\psi} \left[\hat \sigma^{x}_{i},\overline{\hat \sigma^{z}_{j}}(t)\right]^{\dagger} \left[\hat \sigma^{x}_{i}, \overline{\hat \sigma^{z}_{j}}(t) \right]\Ket{\psi}
    \\ & \approx \Bra{\psi} \left[\hat \sigma^{x}_{i},\hat \sigma^{z}_{j}(t)\right]^{\dagger} \left[\hat \sigma^{x}_{i}, \hat \sigma^{z}_{j}(t) \right]\Ket{\psi}
    + \Bra{\psi} \left[\hat \sigma^{x}_{i},\hat \sigma^{z}_{j}(t)\right]^{\dagger} \left[\hat \sigma^{x}_{i}, - i\varepsilon \int_{0}^{t}{\rm d}\tau \left\{\hat \sigma^{z}_{j}(t),\hat H_{\rm ZZ}(\tau)\right\} \right]\Ket{\psi}
        \\ & \qquad+ \Bra{\psi} \left[\hat \sigma^{x}_{i},- i\varepsilon \int_{0}^{t}{\rm d}\tau \left\{\hat \sigma^{z}_{j}(t),\hat H_{\rm ZZ}(\tau)\right\}\right]^{\dagger} \left[\hat \sigma^{x}_{i}, \hat \sigma^{z}_{j}(t) \right]\Ket{\psi}
    \\ & = \comm^{\rm ideal}
         - i\varepsilon \int_{0}^{t}{\rm d}\tau
        \Bra{\psi}\Bigg( \left[\hat \sigma^{x}_{i},\hat \sigma^{z}_{j}(t)\right]^{\dagger} \left[\hat \sigma^{x}_{i},\left\{\hat \sigma^{z}_{j}(t),\hat H_{\rm ZZ}(\tau)\right\} \right] - \left[\hat \sigma^{x}_{i},\left\{\hat \sigma^{z}_{j}(t),\hat H_{\rm ZZ}(\tau)\right\}\right]^{\dagger} \left[\hat \sigma^{x}_{i}, \hat \sigma^{z}_{j}(t) \right]\Bigg)\Ket{\psi}.
    \label{eq:dCdef}
\end{split}\end{equation}

The integral term, proportional to $\varepsilon$, is the experimental error induced by the always-on $ZZ$ coupling. 

We note this error takes the form of a sum over eight similar terms of the form
\begin{equation}\begin{split}
 \Delta\comm = \comm^{\rm meas} - \comm^{\rm ideal} & = 
 i\varepsilon\int_{0}^{t}{\rm d}\tau
     \sum_{\alpha=1}^{4}\Bra{\psi} \left[\hat \sigma^{x}_{i},\hat \sigma^{z}_{j}(t)\right]^{\dagger}\hat{\mathcal U}_{\alpha}\hat H_{\rm ZZ}(\tau)\hat{\mathcal V}_{\alpha} \Ket{\psi} - \Bra{\psi} \hat{\mathcal V}_{\alpha}^{\dagger}\hat H_{\rm ZZ}(\tau)\hat{\mathcal U}_{\alpha}^{\dagger}\left[\hat \sigma^{x}_{i},\hat \sigma^{z}_{j}(t)\right] \Ket{\psi},
\end{split}\end{equation}
where $\hat{\mathcal U}_{\alpha}, \hat{\mathcal V}_{\alpha}$ are unitary operators defined by Eq.~\ref{eq:dCdef} (e.g.~$\hat{\mathcal U}_{2} = \hat \sigma^{x}_{i}, \hat{\mathcal V}_{2} = \hat \sigma^{z}_{j}(t)$, etc.) 
Owing to their different forms and time integration, it is reasonable to treat these terms as uncorrelated errors, and so we find
\begin{equation}\begin{split}
\abs{\Delta\comm}^2 \approx 
 2\abs{\varepsilon}^2 \sum_{\alpha=1}^{4}\abs{\int_{0}^{t}{\rm d}\tau
     \Bra{\psi} \left[\hat \sigma^{x}_{i},\hat \sigma^{z}_{j}(t)\right]^{\dagger}\hat{\mathcal U}_{\alpha}\hat H_{\rm ZZ}(\tau)\hat{\mathcal V}_{\alpha} \Ket{\psi}}^2.
\end{split}\end{equation}
The magnitudes of these terms depend on the likelihood of finding excitations in adjacent lattice sites, averaged over time. As an approximation, we take this to be the infinite-temperature average, or the average over all configurations $\chi$ at a given excitation number $n_{\rm ex}$:
\begin{equation}
    \hat{\mathcal U}_{\alpha}\hat H_{\rm ZZ}(\tau)\hat{\mathcal V}_{\alpha} \to 
    \left[{\binom{9}{n_{\rm ex}}}\right]^{-1}\sum_{\chi}\Bra{n_{\rm ex},\chi} \hat H_{\rm ZZ}\Ket{n_{\rm ex},\chi} \equiv \langle\langle H_{\rm ZZ}\rangle\rangle_{\rm n_{\rm ex}}.
\end{equation}
Note that this approximation is specific to the operators we have used in our commutator.
Here $\Ket{n_{\rm ex},\chi}$ is the state with $n_{\rm ex}$ excitations arranged in configuration $\chi$. Then,
\begin{equation}\begin{split}
\abs{\Delta\comm}^2 & \approx 
 8\abs{\varepsilon}^2 \abs{\int_{0}^{t}{\rm d}\tau\langle\langle H_{\rm ZZ}\rangle\rangle_{\rm n_{\rm ex}}
     \Bra{\psi} \left[\hat \sigma^{x}_{i},\hat \sigma^{z}_{j}(t)\right]^{\dagger}\Ket{\psi}}^2
    \\ &  =\abs{\sqrt{8}\langle\langle H_{\rm ZZ}\rangle\rangle_{\rm n_{\rm ex}}\varepsilon t}^2\abs{
     \Bra{\psi} \left[\hat \sigma^{x}_{i},\hat \sigma^{z}_{j}(t)\right]^{\dagger}\Ket{\psi}}^2
     \\ & \leq\abs{\sqrt{8}\langle\langle H_{\rm ZZ}\rangle\rangle_{\rm n_{\rm ex}}\varepsilon t}^2
     \Bra{\psi}\abs{ \left[\hat \sigma^{x}_{i},\hat \sigma^{z}_{j}(t)\right]}^2\Ket{\psi}
     = \abs{\sqrt{8 \comm}\langle\langle H_{\rm ZZ}\rangle\rangle_{\rm n_{\rm ex}}\varepsilon t}^2.
\end{split}\end{equation}

We finally find
\begin{equation}\begin{split}
\abs{\Delta\comm} \lesssim
    \sqrt{8\comm}\abs{ \langle\langle H_{\rm ZZ}\rangle\rangle_{\rm n_{\rm ex}}} \varepsilon t.
\end{split}\end{equation}
We observe that the error is bounded by the commutator as $\sqrt{\comm}$, and grows linearly in time. We can therefore evaluate the duration of an experiment before the results become unreliable, where $\Delta\comm \approx \sqrt{\comm}$, as a function of the number of particles (excitations) in the lattice. The results are summarized in Tab.~\ref{tab:errorZZ}.
We note that the time scales involved are significantly larger than what is required for our measurements of OTOCs in Fig.~3,~4, where $n_{\rm ex}\leq 3$.

\section{Probing information propagation with OTOC in a 1d chain}

\begin{figure}
\includegraphics{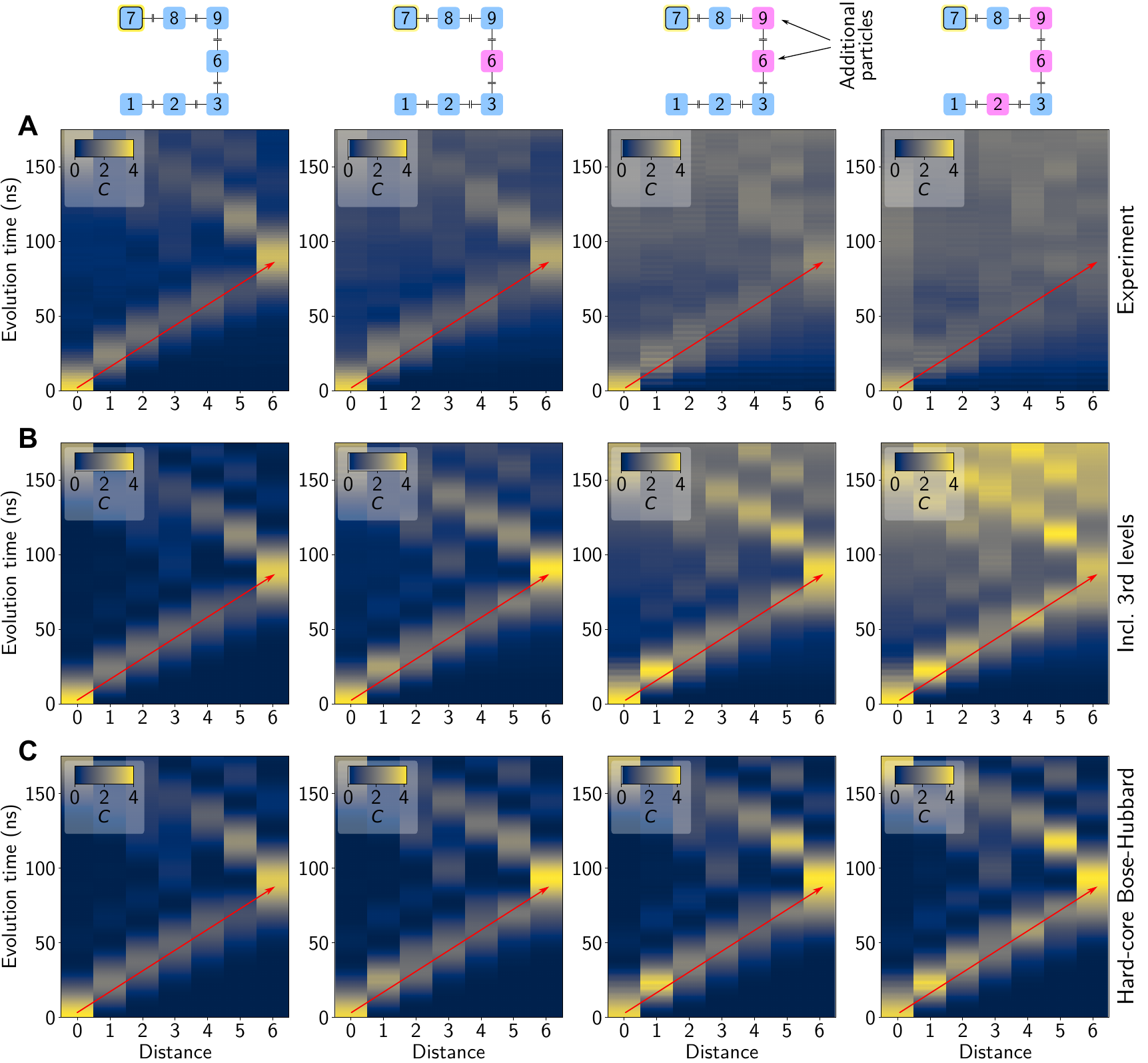}
\caption{\textbf{Probing information propagation in the 1d chain via OTOC measurements}
(\textbf{A}) Experimental data showing the constructed squared commutator $\comm=2+\comm_--\comm_+$ for a varying number of additional particles loaded to the chain.
(\textbf{B}) Numerical simulations based on our experimental Hamiltonian including the third transmon levels. We include the long-range interaction mediated by the two far-detuned qubits that are not part of the 1d chain.
(\textbf{C}) Numerical simulations for the hard-core Bose-Hubbard model.
}
\label{fig:otoc1d}
\end{figure}

We provide a demonstration of the generality and versatility of our technique by probing the propagation of information in a resonant 1d Bose-Hubbard chain of length seven. For this, we are interested in the same OTOC $\otoc$ as in the 2d case (Eq.~8) and we again construct the associated squared commutator $\comm=2+\comm_--\comm_+$. In Fig.~\ref{fig:otoc1d}A we show experimental data that we have obtained for applying the initial perturbation to qubit seven at one of the ends of the chain, and the additional perturbation at time $t$ to sites at a varying distance from qubit seven. We extract data for a different number of additional particles loaded to the lattice and compare with numerical simulations based on our experimental Hamiltonian including the third transmon levels (Fig.~\ref{fig:otoc1d}B) and the ideal simulation of the hard-core Bose-Hubbard model (Fig.~\ref{fig:otoc1d}C). In the realistic simulation (Fig.~\ref{fig:otoc1d}B) we take into account long-range interactions within the chain, mediated by the two qubits that are detuned from the rest of the lattice by $\sim\SI{1.5}{GHz}$. As in the 2d case, the light cone feature survives independent of the number of added particles, illustrating the robustness and generality of this technique for studying the propagation of information.

\begin{figure}
\includegraphics{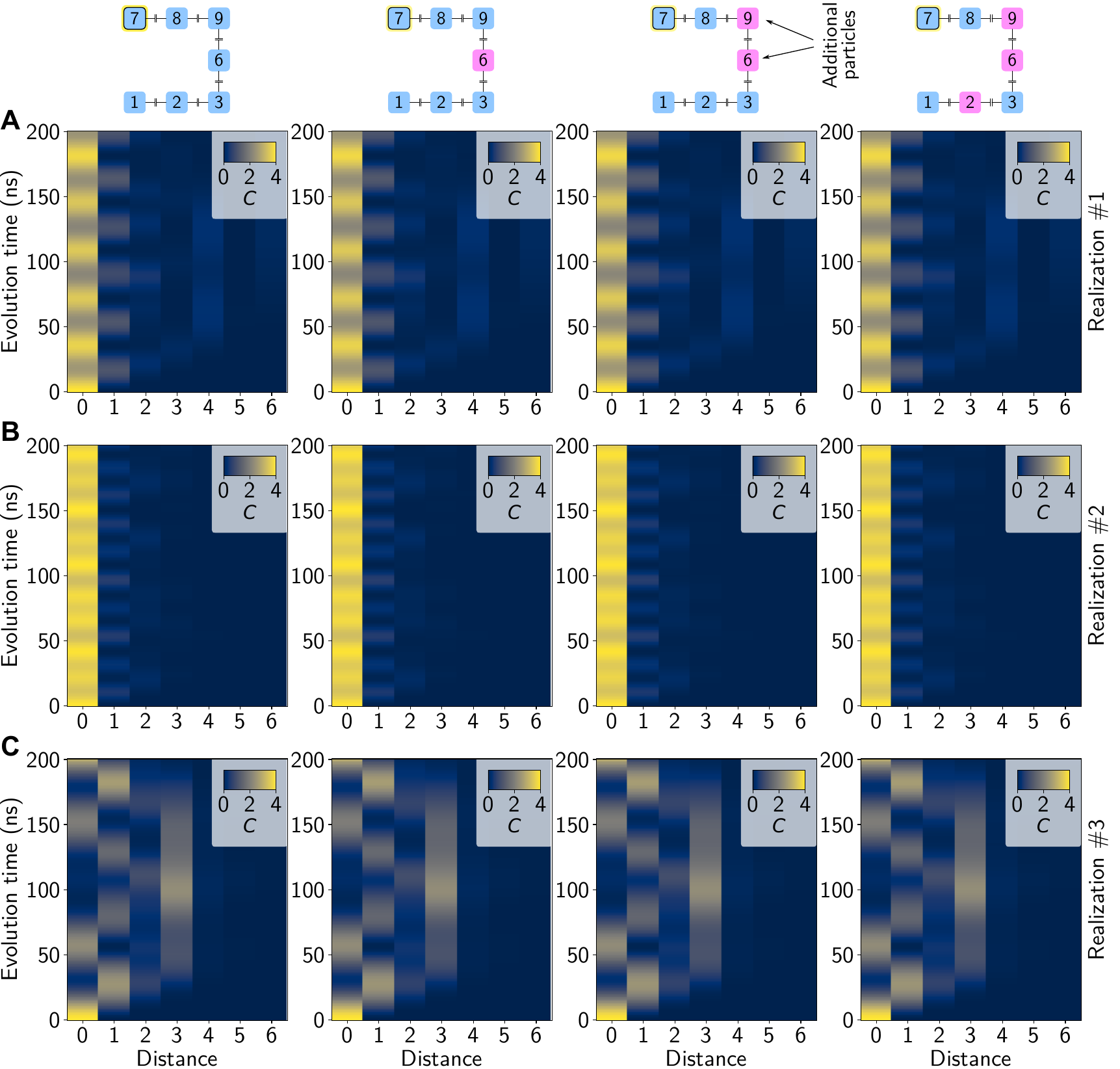}
\caption{\textbf{Information propagation in disordered 1d chains for various number of particles}
Numerical simulations for three random disorder realizations (rows \textbf{A}, \textbf{B}, \textbf{C}) with disorder strength $\langle\langle(\Delta\omega _i)^2\rangle\rangle^{1/2}=2J$ in a seven-site 1d chain. In contrast to the results in Fig.~4, adding more particles to the chain does not help to overcome localization.
}
\label{fig:otoc1ddis}
\end{figure}

The 2d hard-core Bose-Hubbard model is an interacting many-body model without a known mapping to a computationally trivial model. In contrast, the 1d hard-core Bose-Hubbard model maps to free fermions~\cite{Jordan1928}, which is computationally trivial. Consequently, the effect of an `interaction-assisted' overcoming of localization, as observed in Fig.~4, is absent in the 1d hard-core Bose-Hubbard model. This important distinction between the 1d and 2d models underpins the non-trivial, many-body character of the 2d hard-core Bose-Hubbbard model, investigated in our experiment. We verify this via numerical simulations of the information propagation in a seven-site 1d chain subject to random disorder with disorder strength $\langle\langle(\Delta\omega _i)^2\rangle\rangle^{1/2}=2J$. In Fig.~\ref{fig:otoc1ddis}, we show numerical simulation data for three randomly chosen disorder realizations (rows \textbf{A}, \textbf{B}, \textbf{C}), and we observe that the efficiency of information propagation is independent of the number of particles in the lattice, in contrast to our observation in the 2d scenario (Fig.~4).

\bibliography{otoc}